\documentclass[aps, prd, amsmath, amssymb, amsfonts, floats, floatfix, superscriptaddress, nofootinbib, twocolumn, showpacs, showkeys]{revtex4-1}

\usepackage[usenames,dvipsnames]{color}
\usepackage{amsmath,amsfonts,latexsym,graphicx,amssymb,times,appendix,bm,multirow,dcolumn,xspace,verbatim,aas_macros}

\definecolor{webgreen}{rgb}{0,.5,0}
\definecolor{webbrown}{rgb}{.6,0,0}
\usepackage[colorlinks=true,linkcolor=webgreen,filecolor=webbrown,citecolor=webgreen]{hyperref}
\usepackage{subfigure}

\graphicspath{{PaperFigures/}}

\newcommand{\pvec}{\bm{\theta}}
\newcommand{\Msun}{\ensuremath{\mathrm{M}_{\odot}}}
\newcommand{\Mc}{\ensuremath{\mathcal{M}}}
\newcommand{\Mco}{\ensuremath{\Mc_{\rm obs}}}
\newcommand{\Mo}{\ensuremath{M_{\mathrm{obs}}}}
\newcommand{\HM}{\texttt{EOBNRv2HM}}
\newcommand{\TT}{\texttt{EOBNRv2}}

\usepackage[normalem]{ulem}

\begin{document}

\title{The Missing Link: Bayesian Detection and Measurement of Intermediate-Mass Black-Hole Binaries}

\author{Philip B. Graff}
\affiliation{Department of Physics, University of Maryland, College Park, MD 20742, USA}
\affiliation{Gravitational Astrophysics Laboratory, NASA Goddard Space Flight Center, 8800 Greenbelt Rd., Greenbelt, MD 20771, USA}
\affiliation{Joint Space-Science Institute, University of Maryland, College Park, MD 20742, USA}
\author{Alessandra Buonanno}
\affiliation{Max Planck Institute for Gravitational Physics (Albert Einstein Institute), Am M\"uhlenberg 1, Potsdam-Golm, 14476, Germany} 
\affiliation{Department of Physics, University of Maryland, College Park, MD 20742, USA}
\author{B. S. Sathyaprakash}
\affiliation{School of Physics and Astronomy, Cardiff University, Queens Building, CF24 3AA, Cardiff, UK}
\affiliation{Max Planck Institute for Gravitational Physics (Albert Einstein Institute), Am M\"uhlenberg 1, Potsdam-Golm, 14476, Germany} 

\begin{abstract}
We perform Bayesian analysis of gravitational-wave signals from non-spinning, intermediate-mass black-hole binaries (IMBHBs) with observed total mass, $M_{\mathrm{obs}}$, from $50\mathrm{M}_{\odot}$ to $500\mathrm{M}_{\odot}$ and mass ratio $1\mbox{--}4$ using advanced LIGO and Virgo detectors. We employ inspiral-merger-ringdown waveform models based on the effective-one-body formalism and include subleading modes of radiation beyond the leading $(2,2)$ mode. The presence of subleading modes increases signal power for inclined binaries and allows for improved accuracy and precision in measurements of the masses as well as breaking of degeneracies in distance, orientation and polarization. For low total masses, $M_{\mathrm{obs}} \lesssim 50 \mathrm{M}_{\odot}$, for which the inspiral signal dominates, the observed chirp mass $\mathcal{M}_{\rm obs} = M_{\mathrm{obs}}\,\eta^{3/5}$ ($\eta$ being the symmetric mass ratio) is better measured. In contrast, as increasing power comes from merger and  ringdown, we find that the total mass $M_{\mathrm{obs}}$ has  better relative precision than $\mathcal{M}_{\rm obs}$. Indeed, at high $M_{\mathrm{obs}}$ ($\geq 300 \mathrm{M}_{\odot}$), the signal resembles a burst and the measurement thus extracts the dominant frequency of the signal that depends on $M_{\mathrm{obs}}$. Depending on the binary's inclination, at signal-to-noise ratio (SNR) of $12$, uncertainties in $M_{\mathrm{obs}}$ can be as large as $\sim 20 \mbox{--}25\%$ while uncertainties in $\mathcal{M}_{\rm obs}$ are $\sim 50 \mbox{--}60\%$ in binaries with unequal masses (those numbers become $\sim 17\%$ versus $\sim22\%$ in more symmetric mass-ratio binaries). Although large, those uncertainties in $M_{\mathrm{obs}}$ will establish the existence of IMBHs. We find that effective-one-body waveforms with subleading modes are essential to confirm a signal's presence in the data, with calculated Bayesian evidences yielding a false alarm probability below $10^{-5}$ for $\mathrm{SNR}\gtrsim9$ in Gaussian noise. Our results show that gravitational-wave observations can offer a unique tool to observe and understand the formation, evolution and demographics of IMBHs, which are difficult to observe in the electromagnetic window. 

\end{abstract}


\maketitle

\section{Introduction}\label{sec:introduction}

\begin{figure}
\centering
\includegraphics[width=0.98\columnwidth]{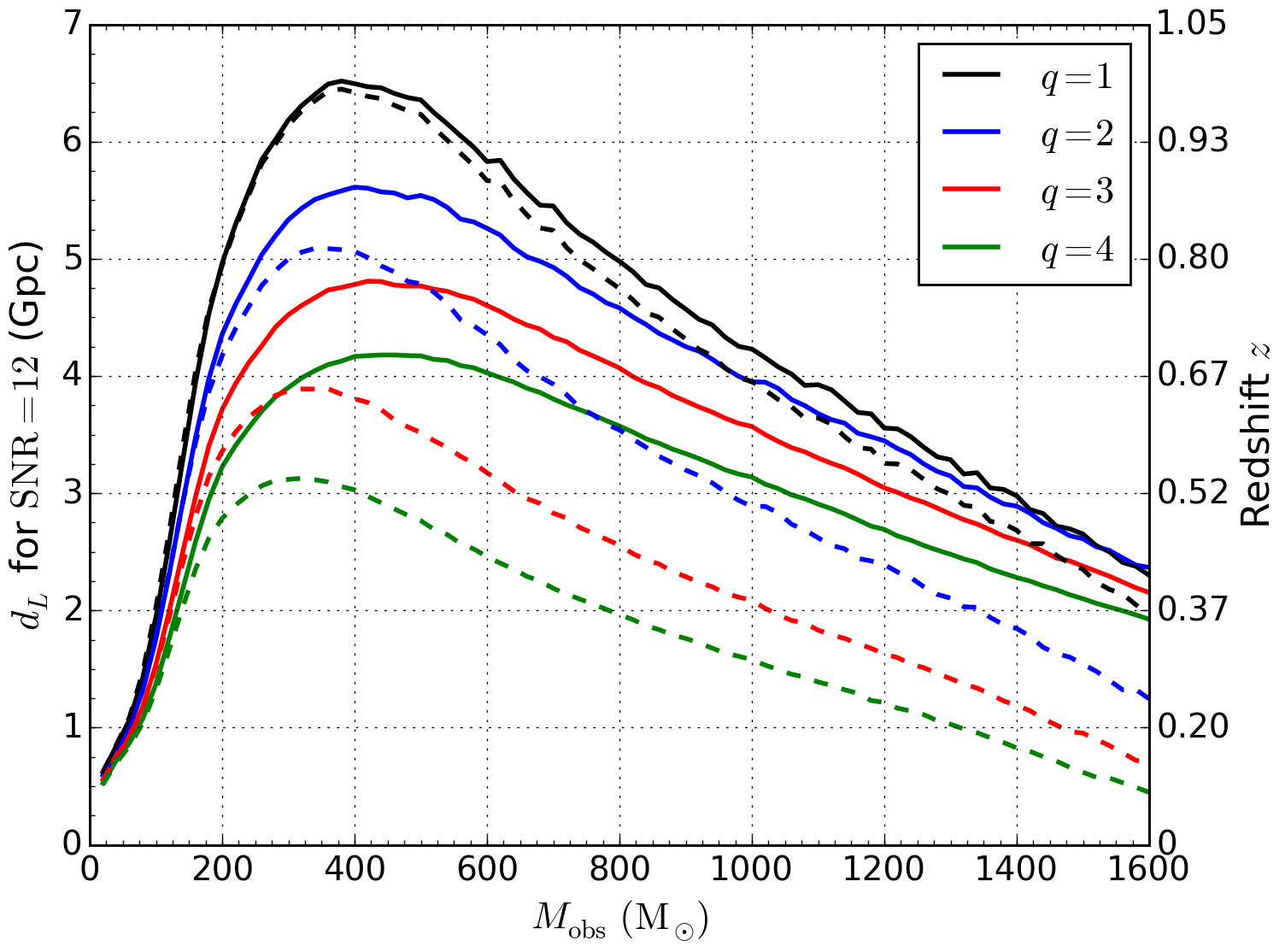}
\caption{We show the distance reach as a function of the observed total mass 
$\Mo$ for several values of the binary mass ratio $q$. The reach is computed 
using a detector network consisting of two advanced LIGO interferometers and 
advanced Virgo, using a network $\mathrm{SNR}=12$.
Detector sensitivities are given by the advanced LIGO and Virgo design curves~\cite{2013arXiv1304.0670L};
the advanced LIGO design is the zero-detuned high-power (ZDHP) noise curve.
The right $y$-axis shows the 
redshift computed assuming cosmological parameters measured by the Planck 
satellite~\cite{Ade:2013zuv}. The continuous curves use inspiral-merger-ringdown 
waveforms with the most dominant five modes (\HM{}), while the dashed 
curve only includes the $(2,2)$ mode (\TT{}).
The values here are numerically averaged over sky location and polarization,
with fixed orbital phase and inclination: $\{\phi,\theta_{\rm JN}\}=\{0,\pi/3\}$ rad.
The coalescence time is also fixed to a GPS time of $t_c = 1000000008\,{\rm s}$, 
corresponding to Sept. 14, 2011 01:46:33 UTC.}
\label{fig:sensitivity}
\end{figure}

Advanced interferometric gravitational-wave (GW) detectors LIGO and Virgo will be turned on in late 2015 (2016 for Virgo) and are
expected to reach design sensitivity by 2019 \cite{Waldman:2011,Aasi:2013wya,2015CQGra..32g4001T,TheVirgo:2014hva}. 
At design sensitivity these detectors will operate in the frequency range from 10 Hz to 1 kHz, with 
an almost flat sensitivity from 40 Hz to 1 kHz. In Fig.~\ref{fig:sensitivity}, we show for the advanced LIGO-Virgo network, 
the distance reach\footnote{For a network consisting of two advanced LIGO and Virgo 
detectors, we compute the distance reach as the root-mean-square distance, averaged over the whole sky
and polarisation angle, at which the network SNR is equal to 12. We do not average over the 
inclination angle, but instead we use a typical value of $\pi/3$.} as a function of observed total mass 
for the full inspiral-merger-ringdown signal of binaries consisting of nonspinning black holes (BHs) and include several multipole modes beyond the 
dominant $(\ell=2, m=2)$ mode, as well as higher order post-Newtonian (PN) corrections (see 
Sec.~\ref{sec:waveforms} for details)~\cite{PhysRevD.84.124052}. 
For nonspinning binary black holes (BBHs) with mass-ratio 1 (mass-ratio 4) of observed total mass 
$\sim 200\Msun$ and $\sim 800\Msun$, the distance reach is $\sim 5$ Gpc (respectively, $\sim 3$ Gpc), with the largest reach of $\sim$ 6.5 Gpc 
(respectively, $\sim 4$ Gpc) for $\sim 400\Msun$ (see Fig.\, \ref{fig:sensitivity}). The intrinsic mass (i.e., the rest-frame mass)  
of a binary $M$ is related to the observed mass $M_{\rm obs}$ by $M_{\rm obs} = (1+z) M,$ and so the 
intrinsic masses detected at these redshifts are significantly smaller than the observed masses.  As we see from Fig.~\ref{fig:sensitivity}, 
subleading modes become increasingly important close to coalescence and their impact on the SNR is relevant for BBHs of total
mass $\gtrsim 200\,\Msun$, especially for asymmetric binaries whose orbital plane is inclined
with respect to the line-of-sight.

The increase in the distance reach brought about by the
use of subleading modes is greater for these latter systems as compared to face-on, 
equal-mass systems where the increase is negligible. 
When spins are included, the distance reach can be a factor of two larger (for near maximal BH spins aligned 
with the orbital angular momentum) or smaller (for maximal spins anti-aligned with the orbital angular 
momentum)~\cite{Buonanno:2005xu,Ajith:2009bn,Belczynski:2014}. Thus, advanced LIGO and Virgo could detect 
BBHs in the hundred solar mass range with a $\mathrm{SNR} = 12$ up to $z\sim 2$, depending on the mass ratio of the system 
and spin.

The above mass range falls in the domain of so-called {\it intermediate mass} black holes (IMBHs). The formation
mechanism, evolutionary history and mass function of IMBHs are largely unknown, as it is very difficult to observe them and measure 
their masses in the electromagnetic window. Several mechanisms have been proposed for their birth and 
evolution~\cite{AmaroSeoane:2007aw,Gair:2010dx,Mandel:2009dn,Belczynski:2014}. 
There is now substantial evidence that galactic nuclei contain massive BHs of millions to 
billions of solar masses but they are believed to have been seeded by lighter BHs of hundreds or
thousands of solar masses (for a review, see, e.g.  Refs.~\cite{Gair:2010dx,Volonteri:2012tp}). While there is 
also firm support for the existence of stellar mass BH candidates \cite{Fender:2012tx}, the IMBH population 
seems to be missing and there is only indirect evidence of their existence.  For example, 
it is suspected that IMBHs could be responsible for ultra-luminous X-ray sources.
While not all such sources are believed to host an IMBH \cite{Bachetti:2014qsa}, some of them do show evidence of BHs 
of tens to hundreds of solar masses. These include a stellar mass BH of $<15\Msun$ in NGC7793 \cite{Motch:2014ska}, 
a more massive $20\Msun \mbox{--} 30\Msun$ BH in M101 ULX-1 \cite{Liu:2013jwd} and a $\sim 400\Msun$ IMBH in M82
\cite{Pasham:2015tca}.  

At present we do not know of any IMBH binaries (IMBHB). However, astrophysical scenarios of 
their formation have been proposed in the literature, which include hierarchical growth of black holes 
at galactic nuclei by accretion of gas, stars and compact objects (i.e. neutron stars and black holes)
and dynamical capture of smaller black holes by nuclear black holes in stellar clusters. 
Hierarchical models of structure formation predict that supermassive BHs found in galactic 
nuclei might initially be IMBHs that grow to their current size by accreting gas and merging 
with other IMBHs~\cite{Volonteri:2012tp,Volonteri:2012ig,Sesana:2009b,AmaroSeoane:2009ui,
AmaroSeoane:2007aw,Gair:2010dx}. In such a scenario we might expect mergers of IMBHBs when 
the Universe began assembling the large structure at high redshift ($z \sim 10\mbox{--}20$).  Such 
mergers might have continued in the local Universe, but it is very difficult to compute merger 
rates as we do not fully understand the initial conditions for IMBHs (mass function of seed 
BHs and their spins), their binaries (orbital parameters at formation and population as a 
function of mass ratio), or the process by which they grow (accretion of gas and merger with 
other BHs). 

Besides growing their mass by dynamical capture in stellar clusters, 
massive BHs may form from the collapse of massive stars
and until recently both observations and theoretical arguments
suggested that stars above $150\Msun$ do not form at non-zero
metallicity. However, recent observations of several stars with
current masses larger than $150\Msun$ in the R136 region of the
Large Magellanic Cloud triggered a re-analysis~\cite{Belczynski:2014}
of the possibility that very massive BHs can have stellar
origin. Ref.~\cite{Belczynski:2014} found that very massive
stellar-origin BHs with mass larger than $100\Msun$ can form only
in low-metallicity environments (i.e., $Z \leq 0.1 \mbox{--} 0.4
Z_\odot$), if the initial mass function extends above $500\Msun$
and pair-instability supernovae do not destroy stars with mass above $500\Msun$.
Moreover, the formation of close massive BH binaries
requires that the very massive stars above $500\Msun$ expand by a
factor of 2 and go through and survive a common envelope phase. 
If these requirements are met, then massive BH binaries are expected 
to have mass ratios of at most a few, spins 
primarily aligned with the orbital angular momentum, and negligible eccentricity 
when they enter the advanced LIGO band~\cite{1989ApJ...343..725Q}.
If the above requirements are not met, then they will have too
wide a separation to coalesce within a Hubble time. However, other phenomena
in dense stellar environments (e.g., cluster binary-single
interactions) and in low-density field populations (e.g, Kozai
mechanism in triple systems) can lower the coalescence time of 
wide massive BH binaries. The investigation carried out in Ref.~\cite{Belczynski:2014} 
concluded that on the order of a few massive BH binaries of stellar-origin 
could be observed by advanced LIGO and Virgo. However, due to astrophysical 
and theoretical uncertainties, the number of detections per year can be as high as
hundreds or as low as zero. 

In this paper we use state-of-the-art waveform models to explore how
well advanced GW detectors can measure the physical parameters of an
IMBHB. Signals from IMBHB coalescences have several important features
that should be incorporated in a study of how GW observations will
help to measure the parameters of such systems.  First, as several
previous studies have already pointed out (see, e.g.,
Ref.~\cite{Buonanno:2014aza} and references therin), in advanced GW
detectors, the plunge, merger and quasi-normal-mode ringdown phases
of evolution contribute significantly to the detectors' distance reach
if the binary has a total mass larger than about $\sim 50\Msun.$
This means that we must use the {\it full} signal, that is not only
the adiabatic inspiral phase, but also the merger and ringdown portions. 
Second, binaries formed in the field will most likely have 
negligible eccentricity~\cite{Peters:1964zz} as they enter
the sensitivity band of advanced detectors and can be assumed to 
trace quasi-circular orbits. For binaries undergoing dynamical 
capture or Kozai mechanism in star clusters, advanced LIGO and Virgo 
might detect mild eccentricities~\cite{Wen:2002km},  if $\Mo 
\sim 10\mbox{--}20 \Msun$. For massive BHs, we expect negligible 
eccentricities when the binary enters the detector band. Indeed, for 
a fixed mass ratio and speed at infinity, the pericenter distance at 
capture is proportional to the total mass~\cite{1989ApJ...343..725Q}. 
Thus, the frequency at capture is inversely proportional to the total mass. 
As a result, larger total masses result in lower capture frequencies and 
thus circularize more by the time the binary gets to a fixed frequency, 
such as 10 Hz. 

In this study we assume our systems to have zero eccentricity. 
Thus the gravitational wave emission in \emph{comparable mass} binaries 
will be dominated by the $(\ell=2,m=2)$ mode at twice the orbital 
frequency, at least until merger. Asymmetric systems with unequal masses, 
nevertheless, emit radiation at other multiples of the orbital frequency or subleading 
modes (see, e.g., Sec.~10.4 in Ref.~\cite{Blanchet:2006}). As shown by several
authors~\cite{Hellings:2003,VanDenBroeck07b,2009PhRvD..79h4032A,2010CQGra..27k4001B,Littenberg:2012uj,Cho:2013efmpehh,OShaughnessy:2014mcmcfm,OShaughnessy:2014bhnshh}
these subdominant modes
can be important in the inspiral phase in improving the accuracy
with which parameters are deduced from GW observations, especially
when the mass ratio of the binary is large. The amplitude of those subleading
modes grow more and more toward
merger~\cite{PhysRevD.84.124052,Pekowsky:2013hmbbh,Varma:2014,Bustillo:2015pnnr}. 
As a consequence, relevant properties of the progenitor binary can be recovered 
from the relative amplitudes of the subdominant
modes excited in the BH
remnant~\cite{Berti:2007imrmp,Berti:2007perd,Hadar:2011nrarrd,Kamaretsos:2011um,Kamaretsos:2012bs,London:2014rqnm} 
and tests of general relativity~\cite{Gossan:2011ha,Meidam:2014jpa} 
can be carried out when those subdominat modes are included during merger and ringdown. 
Moreover, for the purposes of
signal candidate detection in template bank searches,
Ref.~\cite{Capano:2014} showed that when constructing banks for BBH 
searches, the inclusion of subdominant modes yielded improved
sensitivity for systems with $\Mo\gtrsim100\Msun$ and $q\gtrsim4$
(where $q\equiv m_1/m_2\geq1$ is the mass ratio). 
Third, waveforms from nonspinning BH binaries on quasi-circular
orbits are very simple chirp-like signals, with monotonically
increasing frequency and amplitude. However, BH spins can cause
amplitude and phase modulations, so one must ideally include spin
effects in the waveform model, unless the IMBH formation 
scenario strongly suggests nonprecessing or negligible spins~\cite{Belczynski:2014}, 
if the BHs grow their mass through multiple 
mergers in stellar clusters. In this paper, however, we will limit
ourselves to nonspinning BH binaries as waveforms that include both
spin effects and subleading modes are not yet available.

\begin{figure*}[ht]
\centering
\includegraphics[width=0.8\textwidth]{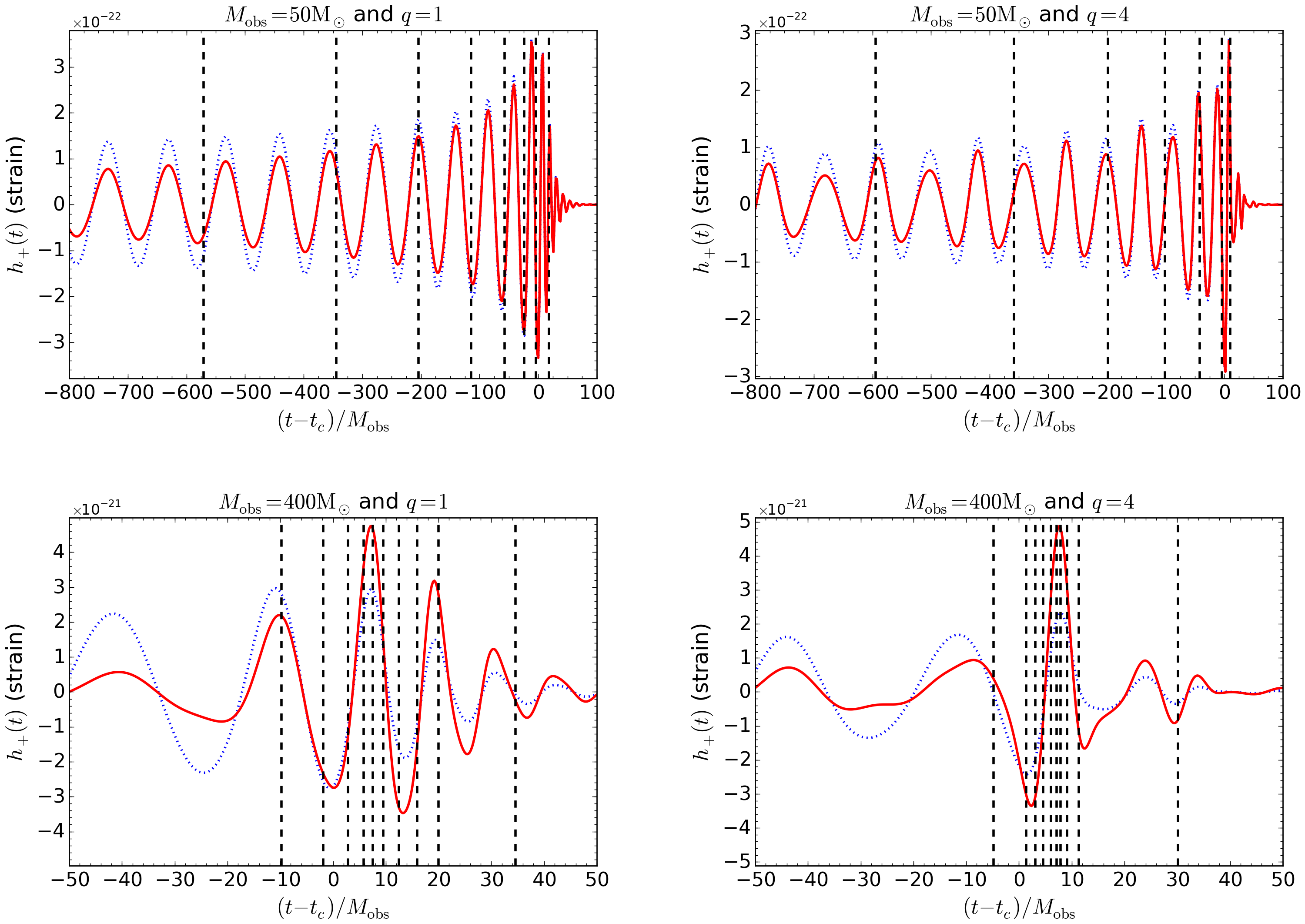}
\caption{We display EOBNR waveforms with subleading modes used in
  this study. We plot the plus polarization of each waveform: original
  (dotted blue), and normalized by the PSD (and then scaled to have
  equal amplitude at $t=0$ as original) (solid red). The plots in the
  top row have $M_{\mathrm{obs}} = 50\Msun$ and in the bottom row have
  $M_{\mathrm{obs}} = 400\Msun$; the left column has $q = 1$ and the
  right column has $q = 4$. The time axis is scaled by the observed
  total mass of the system. Vertical lines are at intervals of $10\%$
  of signal power with the right most line at $99\%$. All systems are
  observed at an inclination of $\theta_{\rm JN}=\pi/3\,\mathrm{rad}$
  and a distance of $d_L=1\,\mathrm{Gpc}$.}
\label{fig:waveform_comp}
\end{figure*}

The rest of the paper is organized as follows. In Sec.~\ref{sec:waveforms} we describe the inspiral, merger and ringdown 
template family used in our analysis, its parameters and the main features introduced by the subdominant modes. 
In Sec.~\ref{sec:bayesinference} we review the basics of Bayesian inference, the sampling technique that we use 
(i.e., nested sampling) and the priors employed in our study. In Sec.~\ref{sec:detection} we discuss how the Bayesian 
evidence of the GW signal can be used to confirm detection, how the false alarm probability can be 
obtained from the Bayesian evidence, and how the Bayesian evidence changes depending 
on the inclusion of the subdominant modes. In Sec.~\ref{sec:measurement} we discuss how Bayesian parameter measurement 
depends on the binary's total mass, mass ratio, inclusion of subleading modes and priors, and compare 
our study to previous ones. We also discuss the astrophysical implications of these measurements of IMBHBs.  Finally, in Sec.~\ref{sec:conclusions} we draw our main conclusions.

\section{Waveforms}
\label{sec:waveforms}

In this section we will discuss the waveform family used in this study
and the parameters used to describe the signal as observed by a detector. In
particular, we discuss the importance of the subleading modes and the merger 
and ringdown phases of the signal for IMBHBs.  We demonstrate this by first 
plotting the signal as observed by an advanced detector and discuss how the SNR is 
accumulated as a function of time. We will also plot the signal power spectrum and
highlight the relevance of subleading modes for unequal-mass systems whose orbital
plane is inclined with respect to the line-of-sight.

\subsection{Inspiral-merger-ringdown waveforms and parameters}

In this study we employ nonspinning waveforms constructed within the effective-one-body (EOB) 
formalism~\cite{Buonanno:1998gg, Buonanno:2000ef} and calibrated to highly-accurate numerical relativity
(NR) simulations having typical length of $30\mbox{--}40$ GW cycles and
mass ratios $q\le6$~\cite{PhysRevD.84.124052}. More specifically,
  we use the {\tt EOBNRv2HM} code in the LIGO Algorithm Library (LAL) \cite{LAL} to 
generate the EOB waveform model in Ref.~\cite{PhysRevD.84.124052},  
which includes four subleading modes, namely the
$(l,m)=(2,1)$, $(3,3)$, $(4,4)$ and $(5,5)$ modes, as well as the
leading $(l,m)=(2,2)$ mode~\footnote{An EOB model with different
  parametrization was subsequently calibrated to the same set of NR
  waveforms used in Ref.~\cite{PhysRevD.84.124052} and provides two subleading modes $(l,m)=(2,1)$ and
  $(3,3)$~\cite{Damour:2012ky}, besides the leading one.}. [These modes come 
from the decomposition of the GW signal $h = h_+ - \imath h_\times$ into $-2$ spin-weighted 
spherical harmonics $_{-2}Y_{lm}$~\cite{PhysRevD.84.124052}.] 
Consequently, GW signals that we study contain the first five harmonics of the orbital frequency. 
During most of the early inspiral phase only the $(2,2)$ mode, and, in particular, its Newtonian
amplitude, will be the dominant component. Other harmonics and PN corrections 
become increasingly important as we get close to merger. The effect of these higher
modes is especially relevant when the binary in question has its merger and ringdown frequencies
in the most sensitive part of a detector's response. The ringdown frequency
of the final remnant of binaries consisting of nonspinning BHs of total mass 
50\,$\Msun$ to 500\,$\Msun$ varies over the range 40 Hz to 400 Hz---the frequency
range where LIGO and Virgo have the best sensitivity, and this provides the motivation for 
our choice of masses used in this study. 

A Markov-chain Monte-Carlo (MCMC)
study~\cite{Littenberg:2012uj} demonstrated that the EOB waveforms of
Ref.~\cite{PhysRevD.84.124052} are indistinguishable from the NR
waveforms~\cite{Buchman:2012dw} used to calibrate them up to $\mathrm{SNR} = 50$
for advanced LIGO detectors. Subsequent investigations carried out in 
Ref.~\cite{Pan:2013tva} verified the accuracy of these nonspinning EOB waveforms
in the entire sensitivity band of advanced LIGO detectors and
suggested that the EOB model be accurate even outside the region of
calibration, i.e. when $q>6$. This expectation was verified by the 
very good agreement found against the $q=10$ NR waveform of $20$ GW cycles in
Ref.~\cite{Hinder:2013oqa} and, especially, against the $q = 7$ NR
waveform of $350$ GW cycles recently produced by the SXS
collaboration~\cite{Szilagyi:2015rwa}.

\begin{figure}[ht]
\centering
\includegraphics[width=\columnwidth]{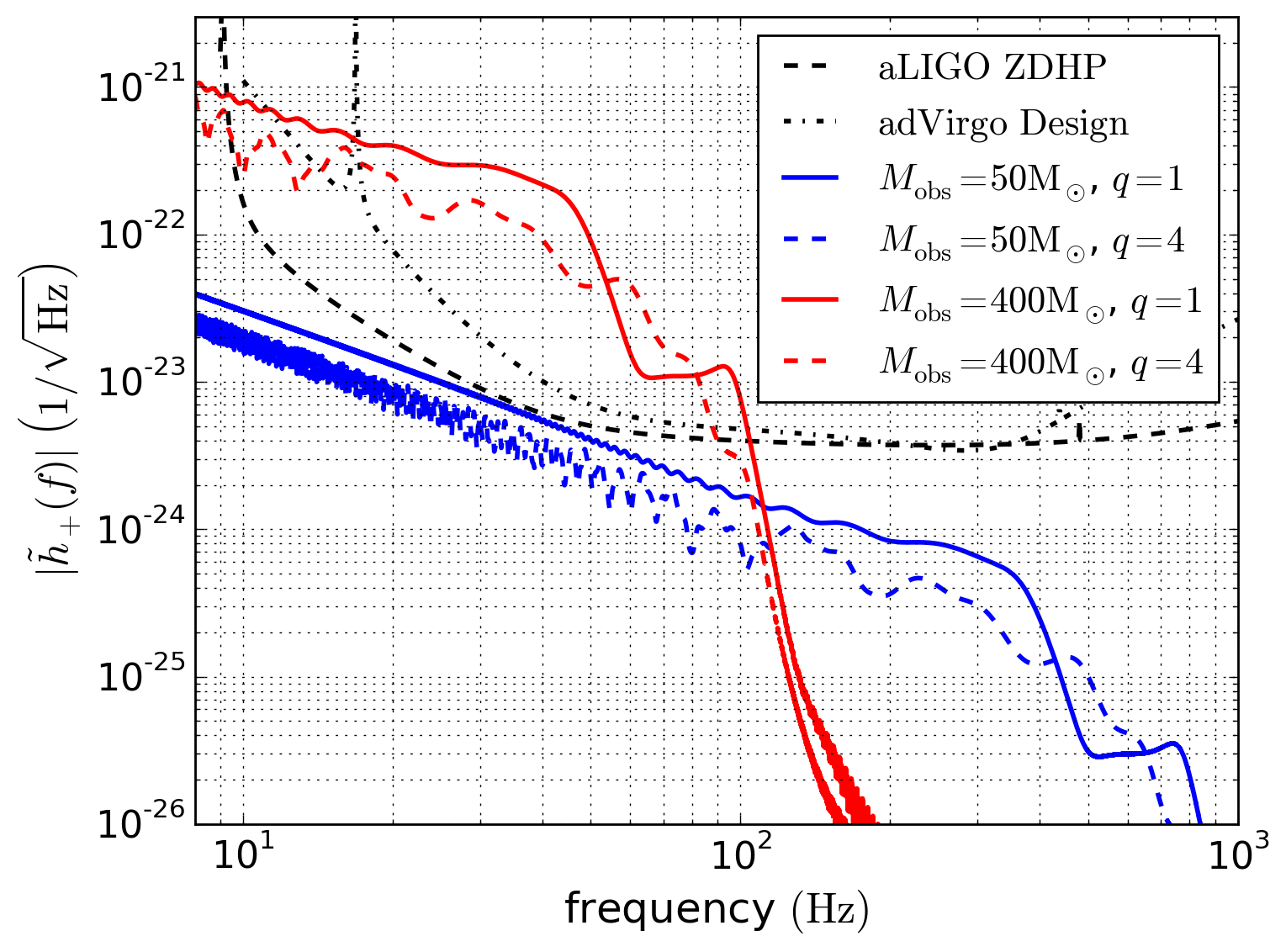}
\caption{We show the Fourier-domain amplitudes of the waveforms displayed in Fig.~\ref{fig:waveform_comp}. 
Extra structure from subdominant modes can be clearly seen in the non-equal mass cases. For comparison, we also 
display the advanced LIGO ZDHP and advanced Virgo design amplitude spectral density 
$\sqrt{S_n(f)}$. For all waveforms we use $d_L=1\,\mathrm{Gpc}$ 
and $\theta_{\rm JN}=\pi/3\,\mathrm{rad}$.}
\label{fig:waveform_psd}
\end{figure}

Since in our study we consider BBHs in quasi-circular orbits with negligible spins, the system can be 
described by nine parameters: $\pvec = \{m_1, m_2, d_L, t_c, \delta, \alpha, \theta_{\rm JN}, \psi, \phi\}$. 
The parameters $m_1$ and $m_2$ are the masses of the
individual BHs. From these quantities we define the total intrinsic mass, $M = (m_1 + m_2)$,  the 
total observed mass, $\Mo = M\,(1+z)$ with $z$ being the redshift, the mass ratio, $q=m_1/m_2 \geq 1$, 
the symmetric mass ratio, $\eta =
m_1 m_2 / (m_1+m_2)^2 = q/(1+q)^2$, the intrinsic chirp mass, $\Mc = M
\eta^{3/5}$ and the observed chirp mass, $\Mc_{\rm obs} = \Mc\, (1+z)$. 
The parameter $d_L$ is the luminosity distance and when combined with
the declination $\delta$ and right ascension $\alpha$, it defines the
sky location of the binary. The time of the peak in the $(2,2)$ mode of
the waveform, as measured at the geocenter, is given by $t_c$; this
serves as an approximation of the merger time. The angle $\theta_{\rm JN}$ measures
the inclination of the binary's total angular momentum
$\mathbf{J}$ (equal to the orbital angular momentum,
$\mathbf{L}$, as the holes are non-spinning) with respect
to the line of sight from the detectors $\mathbf{n}$ 
(geocenter). The polarization $\psi$ and the phase $\phi$ provide the 
additional Euler angles necessary to describe the rotation from $\mathbf{n}$ 
to $\mathbf{J}$.

\subsection{Accumulation of SNR in waveforms}\label{sec:waveformsSNR}

In this paper we are interested in studying the effect on parameter estimation of 
binary systems with larger and larger total masses. When 
we maintain a constant SNR and increase the binary's total mass, 
the signal moves downward in frequency space, resulting in more power from the merger and
ringdown portions as opposed to the inspiral. The merger occurs at
approximately the frequency of the last stable orbit (LSO), which in the
case of a Schwarzschild BH is $f_{\rm LSO} \simeq 4400\,(\Msun/M)
\textrm{Hz}.$ We begin our integration at $f_{\rm min} = 10 \,\textrm{Hz}$,
so for systems with $\Mo \geq 400 \Msun$ there will be little power
from the inspiral portion of the waveform. As the inspiral evolution
is dominated by $\Mc_{\rm obs}$ and the merger and ringdown are
dominated by $\Mo$, we expect the character of the parameter
estimation to transition from one to the other (see Sec.~\ref{sec:measurement}).

To demonstrate this expectation, we compare in Fig.~\ref{fig:waveform_comp} 
two waveforms with $\Mo=50\Msun$ and $400\Msun$. We show both the original 
waveform and the waveform normalized by the advanced LIGO zero-detuned high-power (ZDHP) 
power spectral density (PSD), $S_n(f)$; the normalized waveform has been re-scaled 
so that it has the same amplitude at $t=0$ as the original. The comparisons clearly show that
for the $(\Mo,q)=(50\Msun,1)$ waveform, $90\%$ of the power (SNR$^2$) 
has been accumulated during the inspiral and that the merger and
ringdown play only a minor role.  On the other hand,
for the $(\Mo,q)=(400\Msun,1)$ waveform, only $10\%$ of the power is
collected during the inspiral and now the merger and ringdown are
predominant features. In Fig.~\ref{fig:waveform_psd} we show the amplitudes 
of the waveforms in frequency space in comparison to the advanced LIGO 
ZDHP and advanced Virgo design $\sqrt{S_n(f)}$. We can see in Fig.~\ref{fig:waveform_psd} that
for the higher-mass waveforms, the entire inspiral signal is strongly
down-weighted by the rising of the amplitude spectral density with the merger occurring as 
$\sqrt{S_n(f)}$ reaches a minimum. For the lower-mass waveforms, much of the inspiral
is in the frequency band where the amplitude spectral density is at or near minimum, thereby
allowing this part of the waveform to dominate.

\section{Bayesian Inference}\label{sec:bayesinference}

This section will provide a short background to Bayesian inference. We will focus on the
application of Bayesian analysis to the problems of detection, parameter estimation and model 
selection that will be used in Secs.~\ref{sec:detection} and \ref{sec:measurement} in
the context of GW observations of IMBHBs. After a brief introduction to the basics of Bayesian 
methods, we will discuss a specific technique called {\em nested sampling} that is used to 
efficiently compute Bayesian {\em evidence}, followed by a description of our choice of 
prior probabilities for various parameters and how we compute the {\em likelihood} function.

\subsection{Basics of Bayesian methods}\label{sec:bayesbasics}

Bayesian inference provides a statistically rigorous method of measuring the probability distribution of a set of parameters $\pvec$ given a model or hypothesis $\mathcal{H}$ and a set of data $\mathcal{D}$. Bayes' theorem states that
\begin{equation}
\label{eq:bayes}
\Pr(\pvec|\mathcal{D}, \mathcal{H}) = \frac{\Pr(\mathcal{D}|\pvec,\mathcal{H})\Pr(\pvec|\mathcal{H})}{\Pr(\mathcal{D}|\mathcal{H})},
\end{equation}
where $\Pr(\pvec|\mathcal{D}, \mathcal{H})$ is the posterior probability distribution of parameters used for making inferences about which signal parameters $\pvec$ best fit the data and what the corresponding credible regions are; $\Pr(\mathcal{D}|\pvec,\mathcal{H})$ is the likelihood of obtaining the data given the specific model and parameters, for which we use the shorthand $\mathcal{L}(\pvec)$; $\Pr(\pvec|\mathcal{H})$ is the prior probability of the parameters for the model that represents our knowledge of these values before looking at the data ({\em a priori}); and $\Pr(\mathcal{D}|\mathcal{H})$ is the Bayesian evidence, which is commonly abbreviated as $\mathcal{Z}$. The evidence is the factor needed to normalize the posterior distribution and can therefore also be expressed as
\begin{equation}
\label{eq:evidence}
\mathcal{Z} = \int_{\Theta} \mathcal{L}(\pvec) \Pr(\pvec|\mathcal{H}) d^N\pvec,
\end{equation}
where $N$ is the dimensionality of the parameter space. Since $\mathcal{Z}$ is independent of the parameters, it can be safely ignored for parameter estimation problems, but it is still useful in model comparison.

When comparing two models, $\mathcal{H}_0$ and $\mathcal{H}_1$, one can write their relative probabilities as
\begin{equation}
\label{eq:modelcompare}
\frac{\Pr(\mathcal{H}_1|\mathcal{D})}{\Pr(\mathcal{H}_0|\mathcal{D})} = \frac{\mathcal{Z}_1}{\mathcal{Z}_0} \frac{\Pr(\mathcal{H}_1)}{\Pr(\mathcal{H}_0)},
\end{equation}
where we used Bayes' theorem again and have cancelled out $\Pr(\mathcal{D})$ and substituted in $\mathcal{Z}_i = \Pr(\mathcal{D}|\mathcal{H}_i)$ as appropriate. The relative probability of the two models is thus the ratio of their Bayesian evidences multiplied by the relative probability prior to considering the data. In our analysis, we will take the latter to be $1$ and consider the ratio of the Bayesian evidences, which is called the odds ratio. In the problem of signal detection, $\mathcal{H}_0$ can be considered the noise-only model while $\mathcal{H}_1$ is the signal-plus-noise model. Therefore, an odds ratio much greater than $1$ indicates a strong belief in the presence of a signal. This method naturally incorporates Occam's razor, such that more complicated models are penalized and must sufficiently improve the fit to the data to be favored.

\subsection{Nested sampling and {\sc MultiNest}}\label{sec:nestedsampling}

Nested sampling~\cite{Skilling2006} is a Bayesian inference technique developed for the calculation of the evidence, through which posterior probability samples are produced as a by-product. This is done by transforming the $N$-dimensional integral for $\mathcal{Z}$ into a $1$-dimensional integral over the prior volume. We define the prior volume $X$ by $dX = \Pr(\pvec|\mathcal{H}) d^N\pvec$. We can therefore write the prior probability volume enclosed within a contour (in parameter space) of constant likelihood $\lambda$ as
\begin{equation}
\label{eq:priorvolume}
X(\lambda) = \int_{\mathcal{L}(\pvec)>\lambda} \Pr(\pvec|\mathcal{H}) d^N \pvec.
\end{equation}
The evidence integral of Eq.~\eqref{eq:evidence} can be re-written as
\begin{equation}
\label{eq:evidence2}
\mathcal{Z} = \int_0^1 L(X) dX,
\end{equation}
where $L(X)$ is the inverse of Eq.~\eqref{eq:priorvolume} (returns the likelihood at which a prior volume of $X$ is enclosed) and is a monotonically decreasing function of $X$ (i.e. more prior volume implies lower likelihood contour bound). If we can evaluate likelihood values $L_i = L(X_i)$ such that $X_i$ is a sequence of monotonically decreasing values, the evidence can be computed as a simple sum
\begin{equation}
\label{eq:evidencesum}
\mathcal{Z} = \sum_{i=1}^M L_i w_i \,.
\end{equation}
Here, the $w_i$ are weights which can be taken from a simple trapezium rule such that $w_i = \tfrac{1}{2}(X_{i-1}-X_{i+1})$.

The individual prior weight of each sampled point can also be estimated from the sequence of $X_i$ values. This may be combined with the computed likelihood for that point and the evidence to produce a final posterior probability for the point. The full sequence of points can then be re-sampled accordingly to the points' individual probabilities to produce a set of samples from the posterior.

Nested sampling operates by starting with an initial set of `live' points sampled from the prior distribution. Iterations are then performed whereby the point with lowest likelihood value is removed from the live point set and a new point is sampled from the prior with the restriction that it has higher likelihood than the point just removed. This removal and replacement is continued until a stopping condition is reached (e.g., a tolerance on the evidence calculation). The difficult task here lies in the efficient sampling of new points under this restriction. As the likelihood contour moves upwards, the volume of the prior within that contour will decrease to very small values, making direct sampling of the prior very inefficient. The {\sc MultiNest} algorithm~\cite{Feroz:2008,Feroz:2009,Feroz:2013} addresses this by enclosing the live points in clusters of ellipsoids. A new sample can then be made from the ellipsoids very quickly and as they shrink along with the live points, they create effective likelihood contours to be sampled from, thereby greatly increasing the sampling efficiency. The ellipsoids can be distributed to enclose degenerate and multimodal distributions, making this approach very robust.

{\sc MultiNest} is implemented within BAMBI~\cite{Graff:2012}, which is linked with the \texttt{LALInference}~\cite{2015PhRvD..91d2003V} code of LAL. 
The \texttt{lalinference\_bambi} sampler is used for analysis of simulated signals in this study.

\subsection{Priors used}\label{sec:bayesPriors}
The priors used in this analysis are flat in the component masses, with $\{m_1,m_2\} \in [10, 600]\,\Msun$ and $m_1 \geq m_2$. In Section~\ref{sec:measurementPriors}, we assess the effect of changing the prior by implementing an alternative mass prior that is flat in the log of chirp mass for $\Mc \in [2.45, 435.275]\,\Msun$ and flat over $\eta \in [0.03, 0.25]$.

In both setups, the source location prior is uniform in volume, thus proportional to $d_L^2$ for $d_L \in [100 \,\textrm{Mpc}, 10 \,\textrm{Gpc}]$ and flat in $\sin(\delta)$ and $\alpha$ for $\delta \in [0,\pi]\,\textrm{rad}$ and $\alpha \in [0,2\pi)\,\textrm{rad}$. We use a prior flat in coalescence time that is centered on the true value with $\Delta t_c \in [-0.1,0.1]\,\textrm{s}$. The orientation angles are assumed to be isotropically distributed, thus flat in $\sin(\theta_{\rm JN})$, $\psi$, and $\phi$ for $\theta_{\rm JN} \in [0,\pi]\,\textrm{rad}$, $\psi \in [0,\pi)\,\textrm{rad}$, and $\phi \in [0,2\pi)\,\textrm{rad}$.

\subsection{Likelihood function}\label{sec:bayesLike}

In general, the data obtained from adavanced LIGO and Virgo detectors is the sum of signal, $\mathbf{h}$, and noise, $\mathbf{n}$,
\begin{equation}
\mathbf{d} = \mathbf{h} + \mathbf{n}.
\label{eq:datamodel}
\end{equation}
The signal in a given detector is given by
\begin{equation}
\mathbf{h} = \mathbf{F}_{+}(\alpha,\delta,\psi) \mathbf{h}_{+} + \mathbf{F}_{\times}(\alpha,\delta,\psi) \mathbf{h}_{\times},
\label{eq:signalmodel}
\end{equation}
where $\mathbf{h}_{+,\times}$ are the two independent GW polarizations and $\mathbf{F}_{+,\times}(\alpha,\delta,\psi)$ are the antenna response functions~\cite{2001PhRvD..63d2003A} that depend on the source location and polarization. The antenna response is slowly varying in time due to the rotation of the Earth, but this effect is small for the short duration of the signals considered in this study ($<2$ min).

The noise is modeled as independent and Gaussian in each frequency with a mean of zero and variance given by the detector's PSD. Therefore, the probability of a data stream, $\mathbf{d}_a$, in detector $a$ containing a given signal, $\mathbf{h}(\pvec)$, is given by the probability of the resulting noise realization, $\mathbf{n} = \mathbf{d} - \mathbf{h}(\pvec)$. This is given by the product (sum in log-space) of the probability of the noise for each frequency bin~\cite{2015PhRvD..91d2003V}:
\begin{align}
&\log \mathcal{L}_a(\pvec) = \log \Pr(\mathbf{d}_a \vert H_S, \pvec, S_n(f))= \notag \\
& -\frac{1}{2} \sum_i \left[ \frac{4}{T} \frac{\lvert \tilde{\mathbf{d}}_{a,i} - \tilde{\mathbf{h}}_i(\pvec) \rvert^2}{S_n(f_i)} + \log\left(\frac{\pi T S_n(f_i)}{2}\right) \right],
\label{eq:likelihood}
\end{align}
where $T$ is the segment length, the tilde indicates the discrete Fourier transform of the function, and $i$ is an index over frequency bins. The noise power spectral density $S_n(f)$ will vary from detector to detector and here we use the ones at design 
sensitivity for advanced LIGO and Virgo~\cite{2013arXiv1304.0670L}\footnote{See \url{https://dcc.ligo.org/LIGO-P1200087-v19/public} for PSD data files}. 
$H_s$ indicates that we are using the signal model that assumes a signal is present; this will be compared to the noise-only model, $H_n$, where $\mathbf{h} = \mathbf{0}$. The final likelihood is the 
product of likelihoods from the individual detectors,
\begin{equation}
\log \mathcal{L}(\pvec) = \sum_a \log \mathcal{L}_a(\pvec).
\label{eq:finallike}
\end{equation}
To simulate the sensitivity for advanced LIGO and Virgo detectors, we use a minimum frequency of 
$f_{\textrm{min}} = 10 \,\textrm{Hz}$. In order to include the highest ringdown mode for the lowest possible total mass system, we use a sampling rate of $4096 \,\textrm{Hz}$, giving a Nyquist frequency of $f_{\textrm{Nyq}} = 2048 \,\textrm{Hz}$ for the upper bound of our likelihood sum. A segment length of $128 \,\textrm{s}$ ensures that no waveforms are cutoff in-band.

\begin{figure}
\centering
\includegraphics[width=\columnwidth]{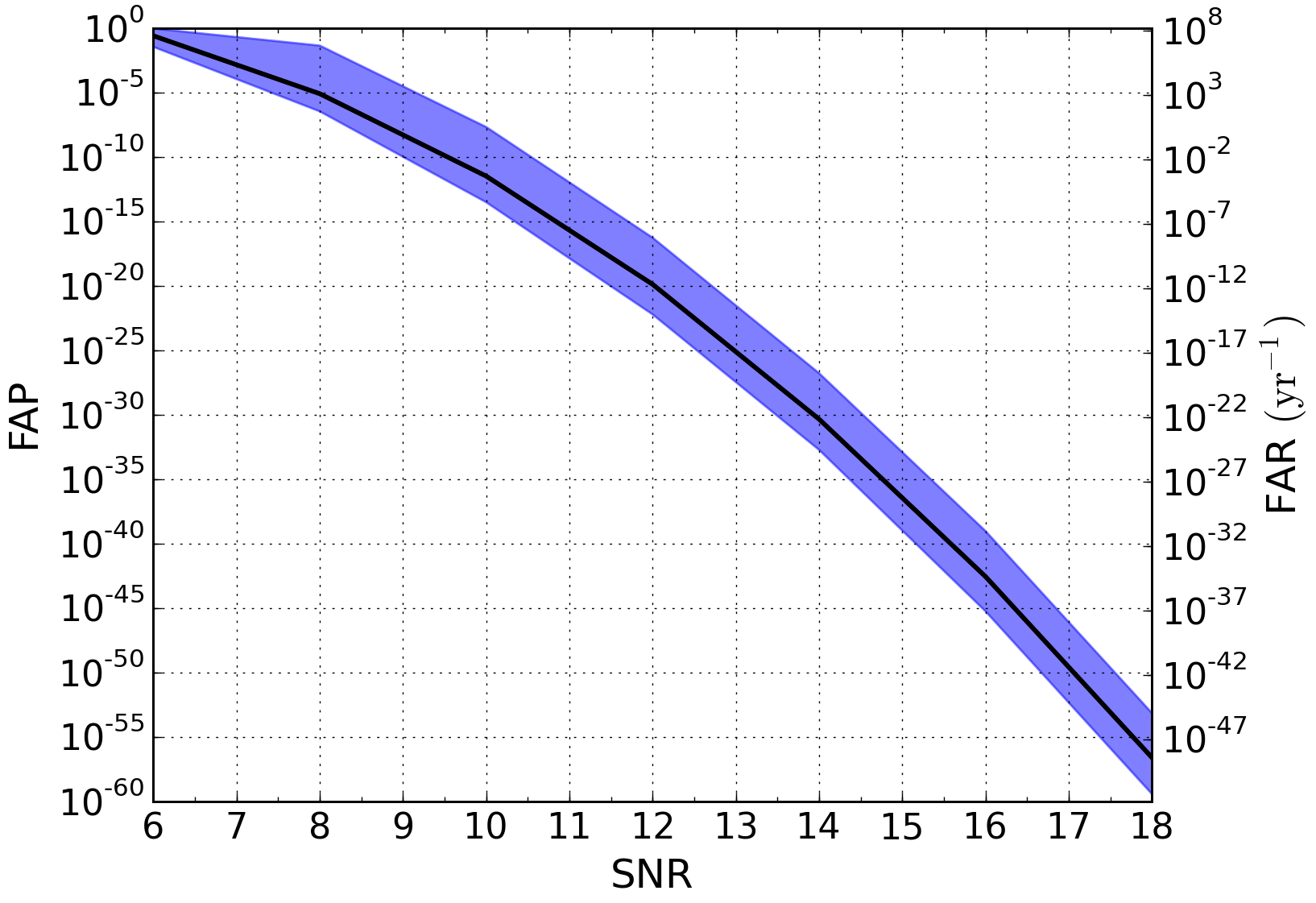}
\caption{False alarm probabilities and rates computed using Bayesian inference for signals buried in Gaussian noise as a function of the injected signal's SNR. The solid line is the median FAP/FAR and the shaded area covers the range from minimum to maximum values of FAP/FAR. Confident $5$-$\sigma$ significant detections can be calimed for $\mathrm{SNR} \gtrsim 9$ but SNRs $\sim 12$ are needed in reality for the same significance in real data that is often non-Gaussian and nonstationary.}
\label{fig:FAPplot}
\end{figure}

\section{Bayesian detection}\label{sec:detection}

It is computationally infeasible to perform a Bayesian analysis
over the entire detector data set, for all signal types, at all times. Therefore,
alternative analysis pipelines are used to first produce candidate triggers for follow-up
analyses using Bayesian inference over a small data set and parameter 
space~\cite{Abadie:2012cbclm,Aasi:2013cbchm}. In the search for BBHs, a discrete bank of template
waveforms~\cite{Sathyaprakash:1991,Balasubramanian:1996prd,Owen:1996prd} is used to
perform matched filter analysis of the data. The matches that cross a pre-set threshold
are ranked by a re-weighted SNR~\cite{Abadie:2012cbclm,Aasi:2013cbchm} and their
significance is measured by comparison to the estimated background (i.e. noise-generated) triggers. 
This is a frequentist method of detection (and significance measurement) and is very useful for
generating triggers which then receive a more detailed follow-up with
Bayesian analysis and other tools. It is the first step in identifying
and confirming a GW signal with LIGO and Virgo. 

In Bayesian inference, one can make claims on the presence of a signal in data by means of model comparison. This is not the same as trigger/candidate finding, but rather looking at the evidence that the candidate is indeed a real GW signal. We compare the signal-plus-noise and noise-only models, $H_s$ and $H_n$, and their respective evidences, $\mathcal{Z}_s$ and $\mathcal{Z}_n$. The probability that random noise would produce an evidence ratio $\mathcal{Z}_s/\mathcal{Z}_n$ is the false alarm probability (FAP). This is given by
\begin{equation}
\label{eq:fap}
\textrm{FAP} = \frac{1}{1+\mathcal{Z}_s/\mathcal{Z}_n}.
\end{equation}
The relative log-evidence ($\log\mathcal{Z}_s - \log\mathcal{Z}_n$) is output by \texttt{LALInference}. The FAP can be converted into a false alarm rate (FAR) by dividing it by the length of the time window~\cite{PhysRevD.83.122005}. This accounts for the amount of time in which we searched for a signal in the data and assumes that all such time intervals are independent.
\begin{equation}
\label{eq:far}
\textrm{FAR} = \frac{1}{1+\mathcal{Z}_s/\mathcal{Z}_n} \times \frac{1}{\Delta t}.
\end{equation}
The time window used in our prior is $\Delta t = 0.2\textrm{s}$.

We compute the Bayesian evidences of the signal plus Gaussian noise model for all simulated signals using injections at  SNRs ranging 
from $6$ to $18$. Total observed masses ranged from $50\Msun$ to $500\Msun$, mass ratios were $q=\{1.25,4\}$, and inclinations were $\theta_{\rm JN}=\{0,\pi/3,\pi/2\}$. In Fig.~\ref{fig:FAPplot} we show the median FAP and FAR calculated over all signals as a function of the SNR. The shaded area covers the range from minimum to maximum computed FAPs. Confident detection can be claimed for a network $\mathrm{SNR} \gtrsim9$, as this corresponds to a detection outside of the $\pm6\mbox{-}\sigma$ region (FAP $<10^{-5}$). This should be taken with a grain of salt, however, as real data will contain non-Gaussian and non-stationary noise features that will need to be addressed with noise modeling~\cite{Cornish:2014bw,Littenberg:2014bl}; SNRs $\sim 12$ are needed in reality for the same significance in real data that is often non-Gaussian and nonstationary.

\begin{figure}
\centering
\includegraphics[width=\columnwidth]{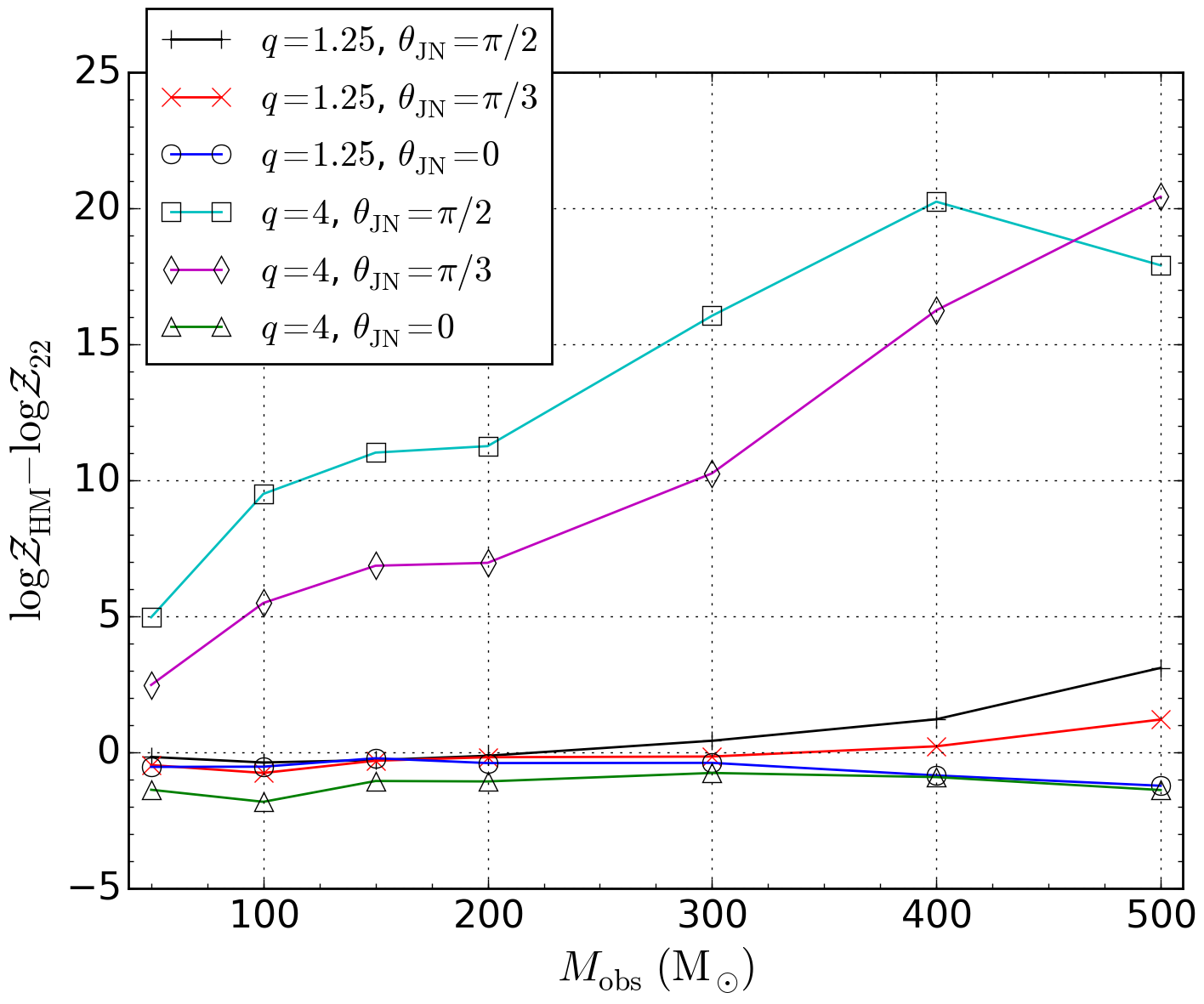}
\caption{We show the difference in log-evidence between using \HM{} and \TT{} templates for recovering 
\HM{} signals.  In cases with significant contribution from subleading modes the \HM{} model is strongly favored.}
\label{fig:BayesLoss}
\end{figure}

In systems where the subleading modes contribute significantly to the
SNR, not including them can result in not recovering the full power of
the signal. This means that using a $(2,2)$-only template (\TT{}) will
yield a lower Bayesian evidence than a more complete template (\HM{})
for the same signal. This loss in evidence will lead to a greatly increased FAP and FAR.
A model comparison between the two will favor the
complete model when these modes are significant -- inclined systems
with larger mass ratios. Fig.~\ref{fig:BayesLoss} shows that in
these cases the \HM{} waveform model including subleading modes will
be strongly favored; when subleading modes contribute little SNR,
neither waveform model is strongly favored over the other (slight
preference for $(2,2)$-only when $\theta_{\rm JN}=0$ and slight
preference for subleading modes when $\theta_{\rm JN}>0$ and $q$ is close
to $1$). 

The importance of including subleading modes in search pipelines was 
investigated in Ref.~\cite{Capano:2014}. They found that 
using waveforms with subdominant modes increases the sensitive region only
for high total masses ($\Mo\gtrsim100\Msun$) and asymmetric
($q\gtrsim4$) IMBHBs. Furthermore, they found that the most
significant gains are in regions of the parameter space with the
lowest expected event rates. Although the study of Ref.~\cite{Capano:2014} 
was limited to component masses $m_i\leq200\Msun$ and $\Mo<360\Msun$, we can expect the
trends to continue for larger masses. The result that
subleading modes are significant in detection only for asymmetric
and large total mass systems is consistent with our findings
described in this section.

\section{Measurement}\label{sec:measurement}

After the detection of a GW signal from a binary system, we perform parameter estimation analysis, 
which involves producing a sufficient number of samples
from the posterior distribution so that we are able to measure peaks
and analyze correlations and degeneracies. Since we expect our first
detections to be at just above threshold, all analyses in this
section -- unless otherwise stated -- use injected signals with a network SNR
of $12$. This is achieved by adjusting the distance of the signal to
obtain this exact value.

In the following sections we discuss our ability to perform parameter
estimation under varying conditions. We estimate the statistical
uncertainty and bias in the measurement of signal parameters; these
are the width of the posterior distribution and the distance between
the peak and the true values, respectively. In creating the data to be
analyzed, no noise realization is added. This eliminates additional
uncertainty and bias introduced by a random noise realization; zero
noise is the most probable realization. This is different from
averaging over many noise realizations, as the latter would result in
increased uncertainty even as the biases cancel out (and would also require many more runs to be performed).

Results presented are predominantly for $q=\{1.25,4\}$ systems. Analyses were also performed where
injected waveforms had $q=\{1,2,3\}$; we found the results to be consistent with those discussed here.
We limited $q\leq4$ for injected signals due to the increased computational cost for higher mass-ratio waveforms.

\begin{figure*}
\centering
\includegraphics[width=\textwidth,height=.75\textheight,keepaspectratio]{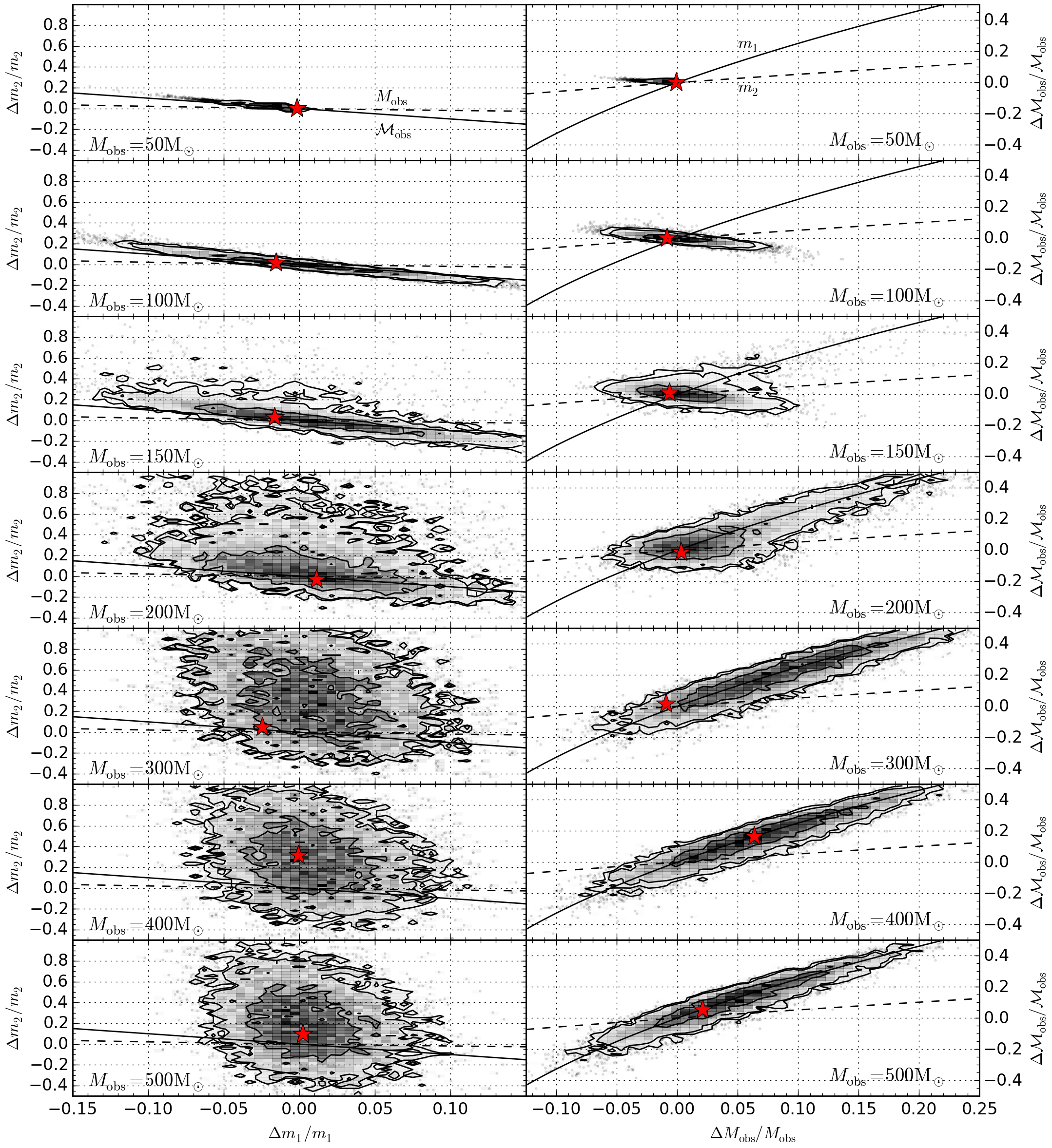}
\caption{Posterior distributions of the mass estimation. All values are presented as fractional errors, i.e., $(x-x_{\textrm{true}})/x_{\textrm{true}}$. The left column displays $m_2$ vs. $m_1$ and the right column displays $\Mc_{\rm obs}$ vs. $\Mo$. The rows are of increasing $\Mo$ from $\Mo=50\Msun$ at the top to $\Mo=500\Msun$ at the bottom. For all systems, $q=4$ ($\eta=0.16$) and $\theta_{\rm JN}=\pi/3$. The star indicates the point with highest $\log\mathcal{L}$ and the contours are at $50\%$, $90\%$, and $95\%$ credible levels (inside to outside). In the left column, the solid lines are of constant $\Mo$ and the dashed lines are constant $\Mc_{\rm obs}$; in the right column, solid is constant $m_1$ and dashed is constant $m_2$. In all cases the lines intersect the true values at $(x,y)=(0,0)$.}
\label{fig:2Dmass}
\end{figure*}
\subsection{Measuring variance with increasing binary's total mass }\label{sec:measurementMtot}

\begin{figure*}
\includegraphics[width=0.4\textwidth]{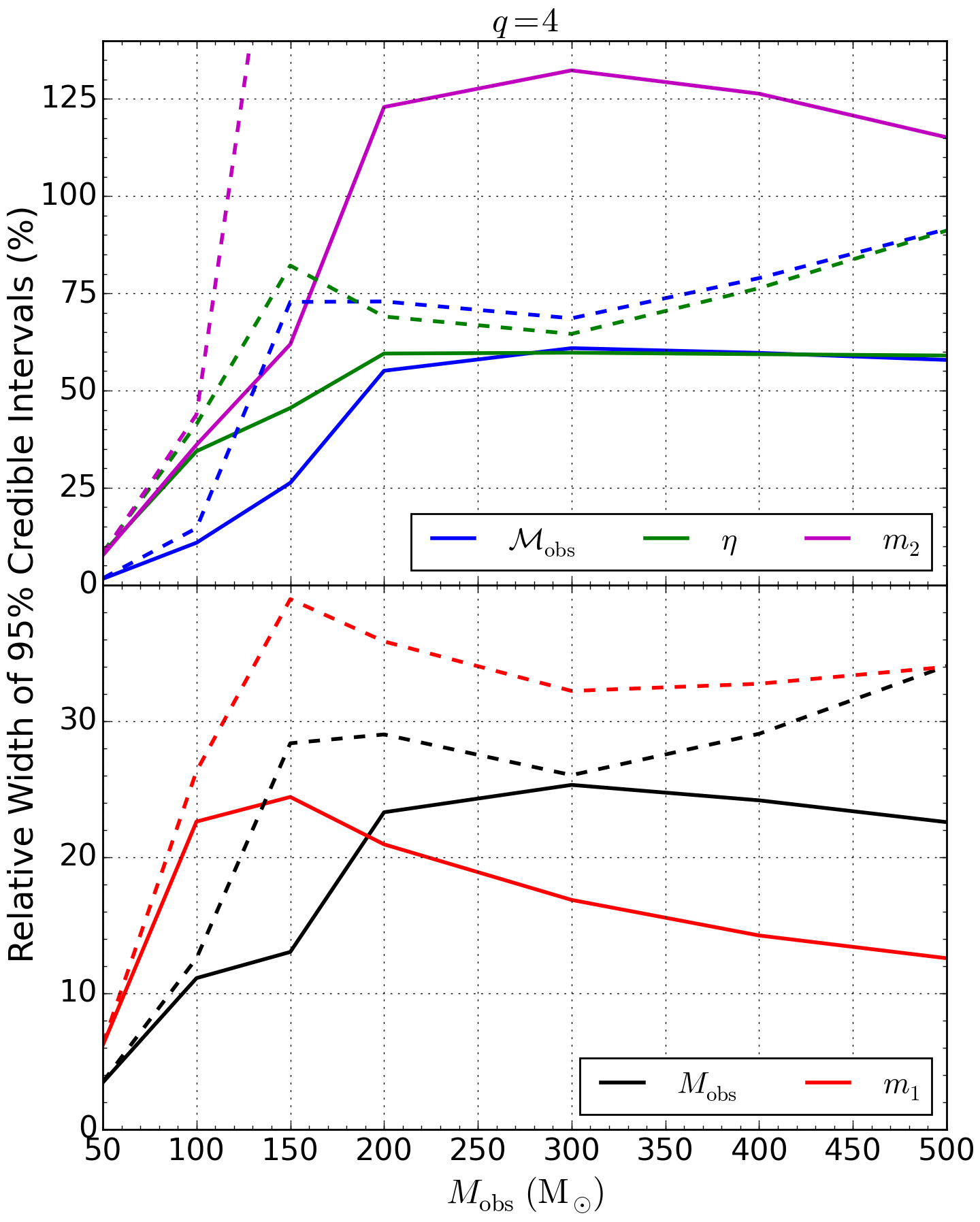}
\hspace{1cm}
\includegraphics[width=0.4\textwidth]{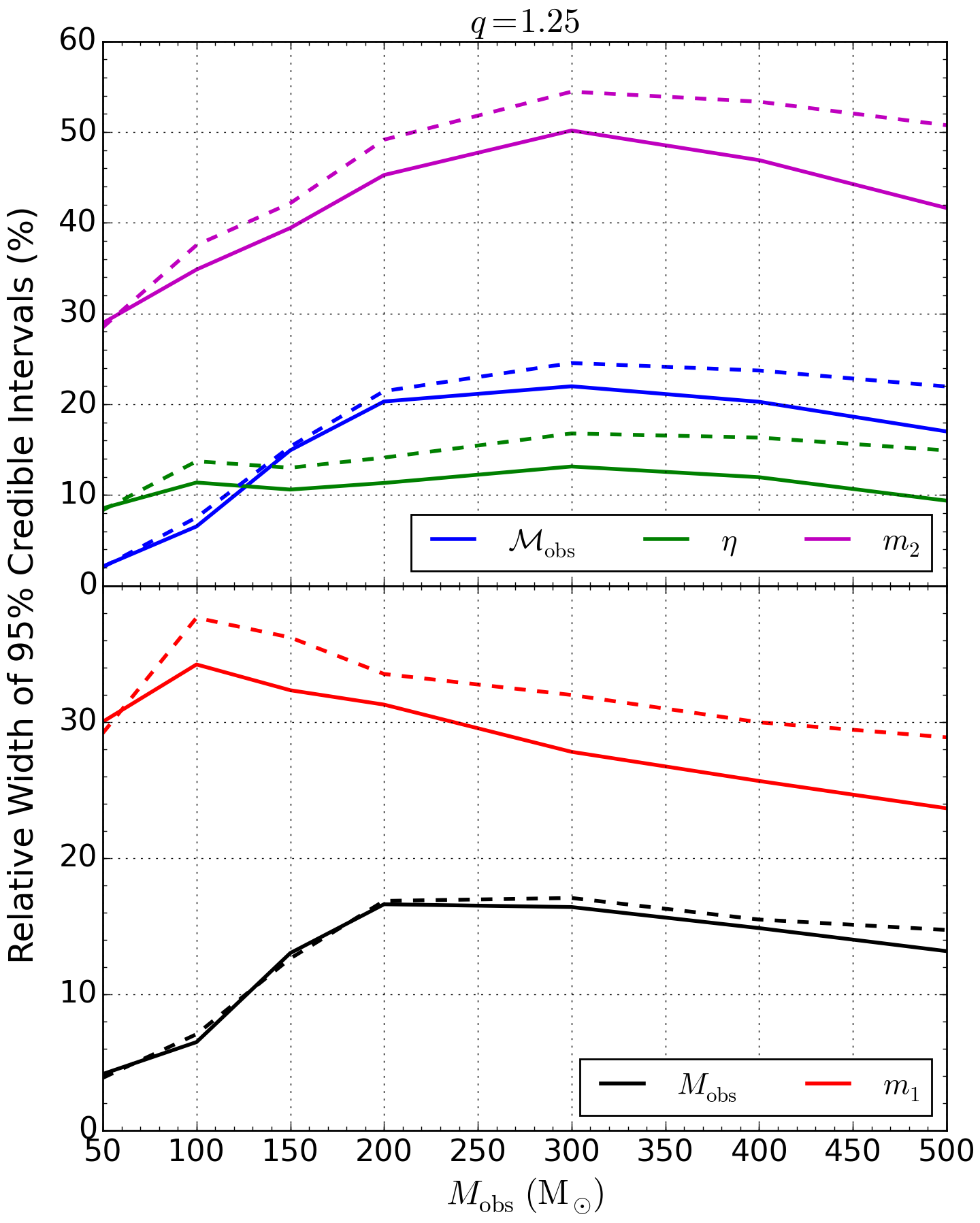}
\caption{Uncertainty in the measurement of different mass parameters: $\Mo$, $\Mc$, and $\eta$. Results here are for a system with (left panel) $q=4$ ($\eta=0.16$) and (right panel) $q=1.25$ ($\eta=0.247$). In both cases, $\theta_{\rm JN}=\pi/3$ and $\mathrm{SNR}=12$. The solid lines are for templates with subleading modes and the dashed lines are for templates with the $(2,2)$ mode only. The relative width is given by $(x_{97.5\%}-x_{2.5\%})/x_{\rm true}$.}
\label{fig:mass_uncertainty}
\end{figure*}

In our first set of comparisons, we study the effect of the total
mass of the system on the estimation of the source's intrinsic
parameters. Specifically, we investigate the statistical errors on the
measurement of $\Mo$, $\Mc$, $\eta$, $m_1$, and $m_2$.
Posterior distributions over $(\Mo,\Mco)$-space and $(m_1,m_2)$-space
are shown for various values of $\Mo$ (always using $q=4$ and $\theta_{\rm JN}=\pi/3$)
in Fig.~\ref{fig:2Dmass}. 
These are summarized in the left panel of Fig.~\ref{fig:mass_uncertainty}, which shows the
relative widths of the $95\%$ ($\pm2\mbox{-}\sigma$) credible intervals
(i.e., $(x_{97.5\%}-x_{2.5\%})/x_{\rm true}$) for the various mass parameters. 
At the lower mass end ($\Mo=50\Msun$), uncertainty is low due to how well the chirp mass 
$\Mc_{\rm obs}$ is measured from the inspiral phase of a waveform. As the total mass increases,
the uncertainty increases and for $\Mo\geq150\Msun$ the uncertainty in
$\Mc_{\rm obs}$ is similar to or greater than that in $\Mo$. This change is due to less
inspiral signal being present in the most sensitive band of the detector; the
ringdown is predominantly dependent on $\Mo$ and therefore this
parameter is measured more accurately. However, the inspiral measures
$\Mc_{\rm obs}$ better than the ringdown measures $\Mo$, so the resulting
uncertainty is larger. Above $\Mo=300\Msun$, the uncertainty decreases
slightly; this is due to the ringdown matching up better with the
minimum of the advanced LIGO/Virgo PSD and the subleading modes moving into
more sensitive regions of the PSD. 

When these same systems are face-on ($\theta_{\rm JN}=0$) or
have lower $q$ (more equal component masses), the uncertainties for the $(2,2)$-only
waveform closely resemble those for the waveform with all modes, just a little larger. This
can be seen in the right panel of Fig.~\ref{fig:mass_uncertainty} and
it is what would be expected for systems with little contribution from the subleading modes. 
In all cases, the templates that include subleading modes of radiation have lower
uncertainty than those using only the $(2,2)$ mode. Additionally, for inclined and 
asymmetric systems, as $\Mo$ increases from
$300 \Msun$ to $500\Msun$, the uncertainty when using $(2,2)$-only
templates grows significantly while that from using templates with
subleading modes slightly decreases. This is due to the fact that the subleading modes
provide information about the mass ratio of the system in their
relative amplitudes and phases. This information contained in the
subdominant modes breaks the model degeneracies and allows us to
better infer the component masses as the ringdown phase of the
waveform enters the most sensitive region of the PSD. These results
can also be seen in Fig.~\ref{fig:2Dmass}. Being able to accurately
measure the component masses is important in allowing us to make
inferences on the source population of these massive BHs~\cite{Belczynski:2014}.

Thus, we observe that at an SNR of $12$ uncertainties for $\Mo$ can reach $\sim20\mbox{--}25\%$ in
asymmetric binaries while uncertainty in $\Mco$ reaches up to $\sim50\mbox{--}60\%$ (these
numbers are $\sim17\%$ versus $\sim22\%$ in more symmetric mass ratio binaries).

In Appendix~\ref{sec:measurementSummary}, we provide summary tables of
relative $95\%$ credible intervals for measurements of the masses, luminosity distance, and
coalescence time. These are given over a range of SNRs for two mass
ratios ($q=\{1.25,4\}$) and several observed total masses at the inclinations
of $\theta_{\rm JN}=\{\pi/3,0\}$.

\subsection{Measuring degeneracies with increasing binary's total mass}\label{sec:measurement2d}

As the IMBHB systems
increase in total mass, there are distinct changes in the two-dimensional
posterior probability distributions. As the inspiral phase evolution
strongly constrains $\Mco$ and $\eta$, in that order, lower mass
systems will have degeneracies that follow contours in these
parameters. With increasing total mass, however, the inspiral becomes
less important and the merger-ringdown part of the signal contributes
significantly or dominantly to the SNR. This is most well described by $\Mo$ with
much weaker dependence on $\eta$. Thus, we expect there to be a change
in the degeneracies present in the mass estimation. The inspiral
dependency on $\Mco$ can be seen in the PN
inspiral waveforms (see Ref.~\cite{lrr-2007-2}); PN approximants are
accurate for early inspiral when the BHs are sufficiently far from
merger. The dependency on $\Mo$ of the ringdown is similarly given by
the quasi-normal mode decomposition derived
in Ref.~\cite{PhysRevD.73.064030} and implemented in the EOB waveform models 
used in our study~\cite{PhysRevD.84.124052}.

This change in the optimally measured parameters as $\Mo$ increases
can be observed in Fig.~\ref{fig:2Dmass}. In the right column, we show 
the posterior distribution of the masses, parameterized as $\Mo$
and $\Mc_{\rm obs}$, over a range of total masses for an asymmetric system
($q=4$, $\eta=0.16$) viewed at an angle ($\theta_{\rm JN}=\pi/3$). In
the top row, $\Mo=50\Msun$ and we can see that the principal
measurement is of the chirp mass $\Mc_{\rm obs}$ -- posterior samples and contours lie along a
line of near-constant $\Mc_{\rm obs}$. As $\Mo$ increases, at $\Mo=150\Msun$ a
second principal direction of degeneracy becomes evident. This is due
to a different combination of the mass parameters becoming
increasingly constrained relative to the others and realizing a new degeneracy in the measurement.

These observations confirm what we see in the one-dimensional
posteriors in Fig.~\ref{fig:mass_uncertainty}.
The chirp mass $\Mc_{\rm obs}$ is initially measured to lower fractional 
error than the total mass $\Mo$; as $\Mo$ increases the uncertainty grows much faster in $\Mc_{\rm obs}$ 
than it does in $\Mo$. The small decrease in uncertainty at the higher
masses is also visible as the contours shrink slightly. We are now
able to see in Fig.~\ref{fig:2Dmass} that this increase in uncertainty is accompanied by a
changing of the dominant degeneracy in the parameters of the waveform
model.

\subsection{Importance of including subleading modes}\label{sec:measurementHH}

\begin{figure*}
\centering
\includegraphics[width=0.9\textwidth]{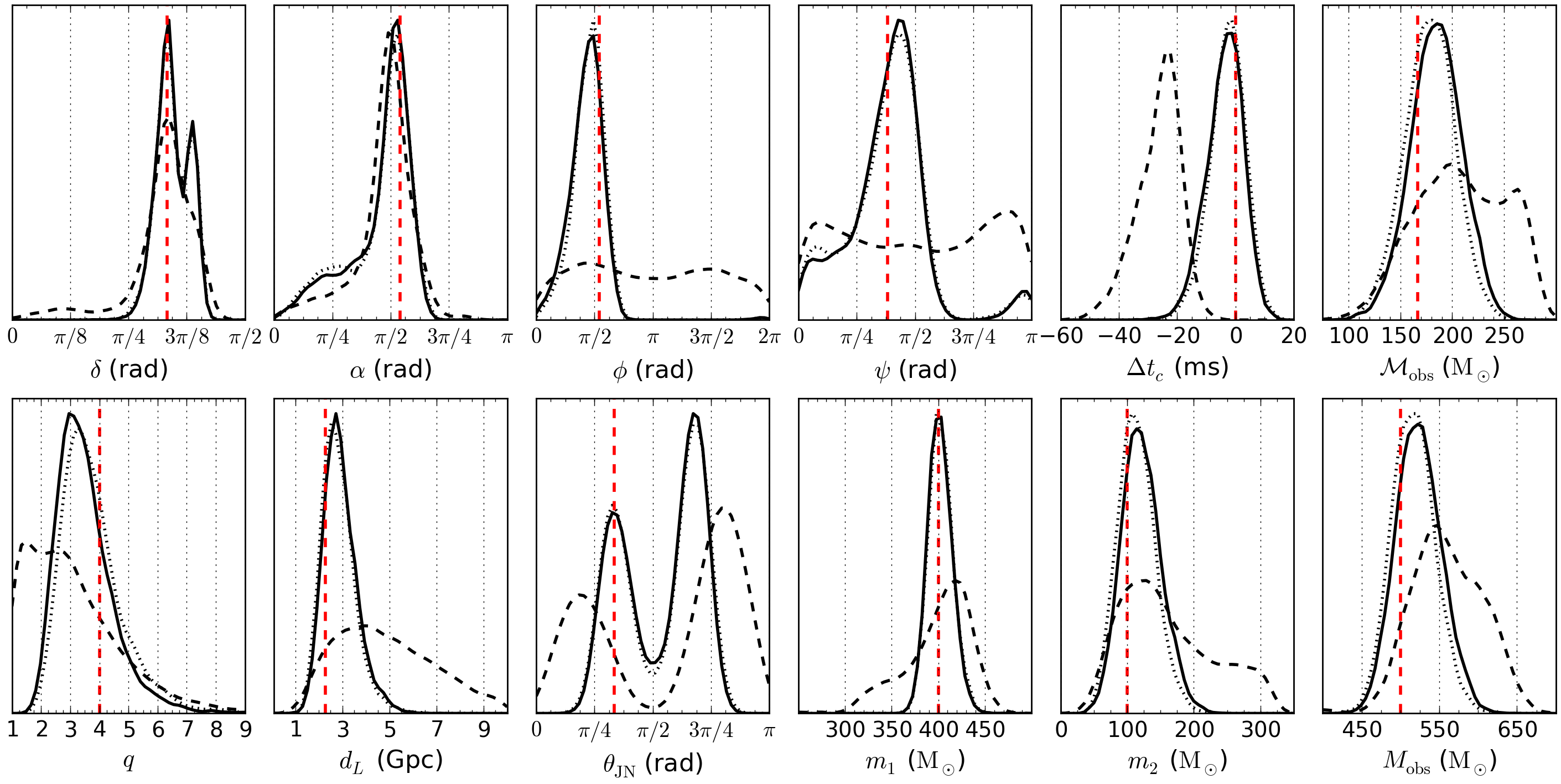}
\caption{One-dimensional posterior distributions for all parameters for an injected signal with $\Mo=500\Msun$, $q=4$, $\theta_{\rm JN}=\pi/3$, and $\mathrm{SNR}=12$. Posteriors are compared for (solid) the \HM{} waveform model which includes subleading modes of radiation, (dashed) the \TT{} waveform model which includes only the $(2,2)$ mode, and (dotted) the \HM{} waveform model along with a different prior distribution that is flat in $(\log(\Mc_{\rm obs}),\eta)$ instead of flat in $(m_1,m_2)$. The vertical dashed red lines indicate the true (injected) value of each parameter.}
\label{fig:HHvs22vsLog}
\end{figure*}

\begin{figure*}[ht]
\centering
\includegraphics[width=0.85\textwidth,keepaspectratio]{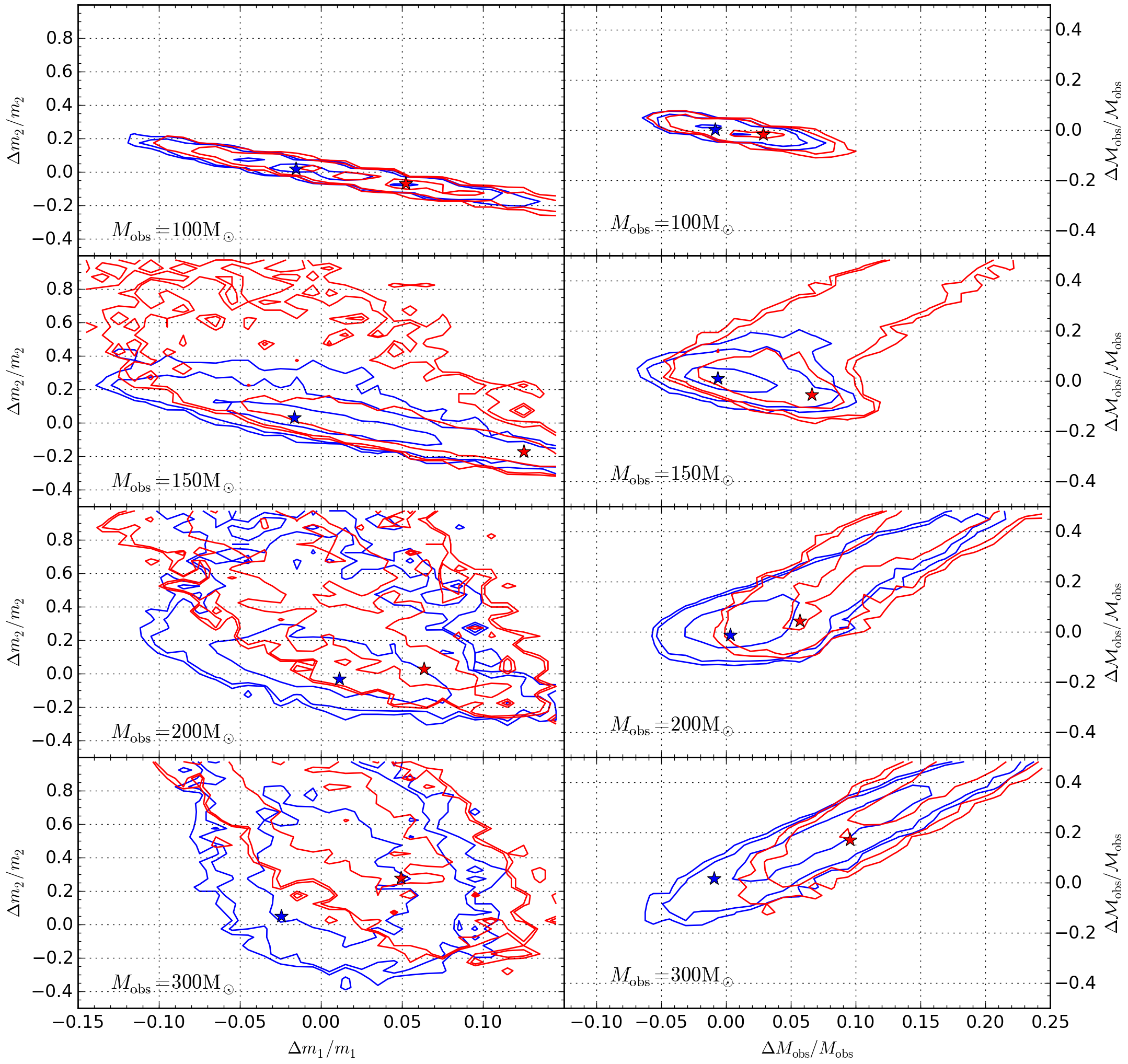}
\caption{Posterior distributions of the mass estimation. All values
  are presented as fractional errors, i.e., $(x-x_{\textrm{true}})/x_{\textrm{true}}$. The left column
  displays $m_2$ vs. $m_1$ and the right column displays $\Mc_{\rm obs}$
  vs. $\Mo$. The rows are of increasing $\Mo$ from $\Mo=100\Msun$ at
  the top to $\Mo=300\Msun$ at the bottom. For all systems, $q=4$
  ($\eta=0.16$) and $\theta_{\rm JN}=\pi/3$. The asterisks indicate
  the point with highest $\log\mathcal{L}$ and the contours are at
  $50\%$, $90\%$, and $95\%$ credible levels (inside to
  outside). Blue contours use \HM{} as a waveform template while
  red contours use \TT{}, which only includes the leading $(2,2)$
  mode.}
\label{fig:2Dmass_HHvs22}
\end{figure*}

As discussed previously, in addition to the leading $(2,2)$ mode, the \HM{} 
waveform model also includes subleading modes $(2,1)$, $(3,3)$, $(4,4)$, and $(5,5)$, 
which introduce additional structure to the waveform and
improve faithfulness to NR waveforms. This increased structure
is important as the relative amplitudes and phasing of the additional
modes introduce information about the source masses. In the ringdown
phase, the additional modes further constrain the mass and spin of the
final BH. This structure creates variation in waveforms
as initial component masses are varied, thereby allowing Bayesian
inference to measure the masses more accurately and precisely as seen
in Figs.~\ref{fig:mass_uncertainty},~\ref{fig:HHvs22vsLog}, and~\ref{fig:2Dmass_HHvs22}.

The modes' structure also contains angular dependence on the
inclination of the system to the detector, $\theta_{\rm JN}$, the
orientation of the orbit in the plane of the sky (polarization),
$\psi$, and the orbital phase of the binary, $\phi$. This structure is
present for the primary $(2,2)$ mode, but the introduction of
additional modes breaks degeneracies in the observed waveform as these
angles vary.

Improvements in measurement of the masses and orientation angles
can all be observed in the example presented in Fig.~\ref{fig:HHvs22vsLog}.
This figure compares one-dimensional
posterior distributions for the case of an injected signal with
$\Mo=500\Msun$, $q=4$, and $\theta_{\rm JN}=\pi/3$. The solid black
line shows the posterior distribution for the waveform template model
including subleading modes (\HM{}) and the dashed line is for the waveform
template model including only the leading $(2,2)$ mode (\TT{}). In both cases,
the signal injected into the data contained the subleading modes -- in
nature, all modes of radiation are  present.

For parameter estimation, the presence of subleading modes 
means that as the angles $\theta_{\rm JN}$, $\psi$, and
$\phi$ vary, there is increased variation of the waveform. This
variation is more prominent for unequal mass binaries and binaries
not observed face-on or face-off ($\theta_{\rm JN} = \{0,\pi\}$), as the
subleading modes will have more significant contributions to the SNR. 
The increased variation allows for more accurate
measurement of $\theta_{\rm JN}$ and breaks degeneracies in $\psi$ and
$\phi$ to allow these two angles to be measured. As $\Mo$ and $\theta_{\rm JN}$
are strongly correlated with the luminosity distance $d_L$ via the amplitude of the waveform,
measuring the former two more accurately means that the latter will be measured more
accurately as well.

The measurement of the coalescence time $t_c$ is offset when using only
the leading mode; this is likely due to slight errors in measuring the
sky position of the source and adjustments in order to align the peak
amplitude of the waveform at merger.

With the additional information provided by the subleading modes in
the \HM{} model, posterior distributions for all mass parameters are
narrower and better centered on the true values. Most notably, the
subleading modes and their relative amplitudes differentiate better
between waveforms with the same total mass, but different mass
ratios. The system injected in the analysis shown in Fig.~\ref{fig:HHvs22vsLog}
has very high $\Mo$ so the merger and ringdown
provide the majority of the SNR. The observed improvement is thus due
to a waveform degeneracy in the mass of the final BH that can be
broken when we are able to measure the final mass and spin of the BH more
precisely. These depend strongly on the mass ratio of the
initial components and are further realized in the relative amplitudes of the
subleading modes. The improved measurement of initial mass values from using
subleading modes can also be seen in Fig.~\ref{fig:2Dmass_HHvs22}, where 
we compare two-dimensional posteriors in the masses between \HM{} and
\TT{} waveform models over a range of $\Mo$ ($q=4$ and $\theta_{\rm JN}=\pi/3$).
Note that when
$\Mo=100\Msun$, the posteriors are nearly identical. However, with
increasing $\Mo$, the late inspiral, merger, and ringdown become
increasingly important. The parameter estimation bias and loss of SNR
is evident when using only the leading mode as the posteriors do not
necessarily peak at or strongly support the true values and the
confidence intervals are considerably larger.

As the observed total mass increases, it is very difficult to measure the mass ratio $q$ (or $\eta$) if the source is face-on. 
Indeed, in this case the subdominant modes are not significant; we find that the posteriors are identical and do not change much with
respect to the true $q$. 
However, when $\theta_{\rm JN}=\pi/3$ or $\pi/2$, because of the presence of the subleading modes, it is possible to measure $q$, 
although the influence of the prior is still evident in tending towards smaller values. In Ref.~\cite{Belczynski:2014},
the authors determine that IMBHBs formed from stellar-origin massive BHs 
will likely have mass ratios $q\leq1.25$. We find that for values of $q\geq2$, in more massive
($\Mo\geq300\Msun$) and inclined systems, we will be able to say that $q>1.25$ with certainty $>90\%$.

In summary, we find that the inclusion of subleading modes of
gravitational radiation improves the accuracy and precision of the
estimation of the source mass parameters as well as some extrinsic
parameters, such as distance and orientation angles. They are significant
for asymmetric and inclined binaries where they contribute more
to the signal's SNR.

\subsection{Effect of priors}\label{sec:measurementPriors}

So far, in all of the analysis runs, we have used a very large prior on the component
masses, which was flat in $(m_1,m_2)$ space. However, one could argue
for other reasonable prior distributions on the masses. One such
alternative is to use a prior that is flat in $\log(\Mc_{\rm obs})$. The quantity 
$\log(\Mc_{\rm obs})$ is used because $\Mc_{\rm obs}$ is a
scaling factor for the waveform amplitude and $\log(\Mc_{\rm obs})$ is the
so-called Jeffreys prior. Additionally, we employ 
a prior that is flat in $\eta$ for the second mass parameter. 

We ran multiple analyses with this second prior option, which is flat in
$(\log(\Mc_{\rm obs}),\eta)$ and find that even at an SNR of $12$ the strength
of the signal is sufficient to render the different prior
distribution a minimal factor. This can be seen in
Fig.\ref{fig:HHvs22vsLog}, where we show the one-dimensional posteriors
from a single analysis. More specifically, we display in dotted lines the 1D posteriors of 
a run with \HM{} waveform model that uses the alternative
prior, to be compared with the solid lines from the run with the
original prior. The lines are nearly identical, with differences much smaller than
those from using the \TT{} waveform; these differences from the alternative prior will continue to
decrease as the SNR is increased.

\subsection{Comparison to previous parameter-estimation work 
with inspiral-merger-ringdown waveforms}\label{sec:measurementAB}

In an earlier work, Ajith and Bose~\cite{2009PhRvD..79h4032A} used 
inspiral-merger-ringdown phenomenological waveform models ({\tt IMRPhenomA}) 
to perform a study similar to ours, but mainly focusing on understanding how 
uncertainties are reduced when merger and ringdown phases are included. Their study 
did not include subleading modes. They examined the statistical error in parameter estimation as given 
by the Fisher information matrix\footnote{The Fisher information matrix measures covariances
  analytically. The square root of the diagonal elements of the
  inverse of the Fisher matrix gives a lower bound on the standard deviation of
  the posterior for the parameters. In the limit of large SNR, this
  estimate becomes exact.} and MCMC analyses. Our uncertainties calculated 
using \TT{} templates ($(2,2)$ mode only) are a factor of a few larger than
those given in Table 1 of Ref.~\cite{2009PhRvD..79h4032A} for
$\textrm{SNR}=10$, $\Mo=\{100,200\}\Msun$, and
$\eta=\{0.25,0.16\}$ (e.g., Ref.~\cite{2009PhRvD..79h4032A} quotes relative uncertainties 
of $2.39\%$ and $3.57\%$ for $\Mo$ in cases with $q=1$ and $\Mo=100\Msun$ or $\Mo=200\Msun$; 
we measure uncertainties of $4.01\%$ and $9.66\%$ in these same cases). The discrepancy in uncertainties is partly
a result of the fact that in Ref.~\cite{2009PhRvD..79h4032A}, the authors maximize their likelihood
function over $t_c$ and $\phi$ and only perform a single-detector search
using an ``effective'' distance that folds in sky position and binary inclination
effects; both of these choices have the effect of fixing values for parameters that we allow
to vary in our more general analysis, thus introducing additional uncertainty. Furthermore, we use \HM{} waveforms as injections, 
as these most closely model NR waveforms, and our \TT{} templates differ from {\tt IMRPhenomA} ones, 
the former being more faithful to NR waveforms~\cite{Hinder:2013oqa}. Despite those differences, the quantitative measurements are in
general agreement and we also find agreement in the qualitative aspects of
increasing uncertainty in mass parameters with increasing $\Mo$. We report our own 
estimations of the uncertainty in Appendix~\ref{sec:measurementSummary}, using the full \HM{} waveform model.

In Ref.~\cite{Littenberg:2012uj}, Littenberg \textit{et al.} examined 
systematic and statistical errors of EOBNR waveforms to assess whether 
those waveforms are indistinguishable from the NR waveforms used to calibrate 
them. The authors employed both \TT{} and \HM{} templates to recover
waveforms generated by NR simulations and, when using subleading modes,
find systematic errors to be comparable to or less than statistical errors for
mass ratios up to $q=6$ and SNRs up to $50$. We find 
statistical errors comparable to the ones of Ref.~\cite{Littenberg:2012uj} for analyses run in common. 
Our results on the importance of subleading modes for unequal-mass and inclined
($\theta_{\rm JN}>0$) systems reaffirm their findings.

Varma \textit{et al.}~\cite{Varma:2014} built on Refs.~\cite{Littenberg:2012uj,Capano:2014}. 
They used as targets ``hybrid'' waveforms constructed by attaching
PN inspiral/EOB waveforms to NR merger-ringdown waveforms, and as templates 
\TT{} waveforms. Instead of examining only
a few points in parameter space, they ran many simulations in order to
average over relative orientation angles. Statistical errors were
computed with the Fisher information matrix. Confirming previous work, they found that
subleading modes are more important for parameter estimation when $\Mo\geq150\Msun$ and $q\geq4$
and important for detection when $\Mo\geq100\Msun$ and $q\geq6$ (see Fig. 1 of Ref.~\cite{Varma:2014}). 
In contrast, our paper employs the full \HM{} waveforms and uses Bayesian analysis, and it 
extends the study to higher $\Mo$.

Bose \textit{et al.}~\cite{2010CQGra..27k4001B} focused on the
importance of including merger and ringdown phases of the
waveform. They analyzed the recovery of inspiral-merger-ringdown  waveforms with
PN inspiral waveforms and found that at masses as low as $\Mo=50\Msun$ 
there are serious systematic errors and an increase in statistical errors due to
the loss in SNR. This result is consistent with our analysis showing the significant amount
of signal power present in the merger and ringdown phases of the waveform
at $\Mo=50\Msun$ and above.

\subsection{Astrophysical implications}\label{sec:astrophysinterp}

As mentioned in Sec.\,\ref{sec:introduction}, the merger rate of IMBHBs is currently highly
uncertain. Consequently, detection of a single event will 
immediately confirm the existence of these systems and constrain their rate. 
In the absence of detection the upper limits reached could be used to rule out some of the models.

As shown in Fig.~\ref{fig:sensitivity}, ground-based detectors will have the
greatest distance reach for equal-mass IMBHBs of observed total mass of 
$\sim 400\Msun$.  The distance reach for such systems will be 
$\sim 6.5$ Gpc or $z\simeq1$.  We find that the comoving volume averaged over all source 
orientations and weighted by the antenna pattern functions of the advanced LIGO-Virgo network 
is $\sim 150\, \mbox{Gpc}^3,$ larger by a factor $\sim 1.8$ than that in Ref.~\cite{Belczynski:2014}.  
This difference can be explained because, as opposed to Ref.~\cite{Belczynski:2014}, we 
consider a detector network (which increases the reach) and use a different SNR 
(which decreases the reach). After five years of non-observation of IMBHBs a rate upper limit of 
$4\times 10^{-11}\,\rm Gpc^{-3}\,yr^{-1}$ can be achieved, which is smaller than 
the rates for most formation models discussed in Ref.~\cite{AmaroSeoane:2007aw,Gair:2010dx,Belczynski:2014}.  
For a binary of same total mass, $\sim 400\Msun$, but mass ratio $q=4$, the reach when including 
subleading modes is smaller by a factor of $1.5$ (see Fig. ~\ref{fig:sensitivity}) and 
the upper limit will be larger by a factor $3.375$, i.e.,  $1.35 \times 10^{-10}\,\rm Gpc^{-3}\,yr^{-1}.$  
Neglecting the subleading modes worsens the upper limit by a factor 2.4. However, Ref.~\cite{Capano:2014} 
showed that in a realistic search the improvement when including subleading modes is significant 
only for mass ratios larger than $\sim 4$.

An important question in cosmology is the mass function of IMBHs. Routine detection of IMBHs 
will help us measure the mass function of component BHs that form merging binaries
and this should be a proxy for the mass function of IMBHs in the Universe, unless 
IMBHBs are formed selectively from a sub-population of IMBHs.  
The component masses of a binary system are strongly correlated and it is not
possible to measure them accurately while using only the dominant mode; subleading modes 
break this degeneracy, especially in the case of asymmetric binaries for which
the mass ratio $q$ is large, helping us measure the component masses more accurately. 
In particular (see Fig.~\ref{fig:mass_uncertainty}), the $95\%$ credible interval in the measurement of 
the heavier companion can be $10\%$ to $25\%$, while the lighter component is measured within $10\%$ to $125\%$, 
depending on the total mass of the binary.  These results are far better than what might be 
possible by electromagnetic observations of such binary systems. Therefore, advanced detectors 
provide the most robust way of determining the mass function across the range of masses from $50\Msun$ to $500\Msun$. 
A related question is the mass function of IMBHBs. Referring to Fig.~\ref{fig:mass_uncertainty}, the total mass is determined to within
a few percent in the case of lighter binaries of $50\Msun$ to within $15\mbox{--}25\%$ for the heaviest 
systems of $500\Msun$ that we consider.  Thus, advanced detectors should help determine the 
mass function of IMBHBs.

\section{Conclusions}\label{sec:conclusions}

In this paper we used state-of-the-art waveform models for inspiral-merger-ringdown phases of evolution 
to estimate uncertainties in parameters of IMBHBs with total mass $\Mo = 50\Msun\mbox{--} 500\Msun$ 
and mass ratio $q = 1\mbox{--}4$. Because for these systems the majority of the SNR is accumulated 
during the last stages of inspiral, merger and ringdown phases, where subleading modes can become 
comparable to the leading $(2,2)$ mode, we also included in the analysis four subleading modes, 
i.e., $(2,1)$, $(3,3)$, $(4,4)$ and $(5,5)$. In particular, we employed the \TT{} and \HM{} waveform 
models in LAL. 

Using a Bayesian analysis, we found that for unequal-mass systems and inclined binaries subleading modes 
improve the measurement of the mass parameters and break degenaracies in distance and orientation 
angles (see Sec.~\ref{sec:measurementHH}). As the binary's total mass increases, the merger and ringdown phases dominate the SNR.  
Since for such high-mass binaries the signal resembles a burst, the measurement will extract the dominant frequency 
of oscillation of the signal, which depends primarily on the total mass, thus the uncertainty in 
total mass becomes smaller than the uncertainty in the chirp mass (see Sec.~\ref{sec:measurementMtot}, 
Fig.~\ref{fig:mass_uncertainty} and Tables~\ref{tab:measurement_uncertainty_q125} and~\ref{tab:measurement_uncertainty_q4}).
In contrast, for lower total masses, the SNR accumulates 
over many cycles of inspiral and the chirp mass is better measured. The presence of subleading modes is less crucial 
for comparable-mass systems and face-on binaries (see Sec.~\ref{sec:measurementMtot}, Fig.~\ref{fig:mass_uncertainty} and Table~\ref{tab:measurement_uncertainty_q125}). Inclusion of subleading modes allows for improved measurement of the mass ratio for asymmetric and inclined systems (see Table~\ref{tab:measurement_uncertainty_q4}).
Finally, as discussed in Sec.~\ref{sec:astrophysinterp}, GW observations of IMBHBs will demonstrate 
the existence and shed light on the demographics of IMBHs, even if component masses will be measured only with a fractional error of
(several) tens of percent. 

Our analysis was restricted to nonspinning IMBHs and explored only part of the parameter space. 
These limitations were a consequence of the fact that EOBNR waveform models are expensive to 
generate for Bayesian analyses. Higher-mass binaries ($\Mo\sim400\Msun$) take tens to hundreds of milliseconds to
generate, while lower-mass binaries ($\Mo\sim50\Msun$)
will require up to tens of seconds; computational time will quickly add up as $10^6\mbox{--}10^7$ 
waveform computations are required for a complete analysis. This cost is compounded by the requirements both for long segments to enclose the entire waveform while in-band
(long due to the low minimum frequency) and a high sampling rate in order to include the subleading modes in the
ringdown signal. Recently, reduced-order models (ROM) have been developed for 
EOBNR waveforms, either for spinning, nonprecessing systems~\cite{2014CQGra..31s5010P} or nonspinning, but with subleading-mode 
waveforms~\cite{Marsat:2014}. Future investigations could employ these faster template families. 

\vspace{0.5truecm}
While completing this work, we became aware of the study of Ref.~\cite{Veitch:2015aa}, which 
includes the effect of nonprecessing spins using a ROM built on the {\tt SEOBNRv2} template 
family~\cite{Taracchini2014,2014CQGra..31s5010P}, while discarding the subdominant modes. 

\acknowledgments

We thank Cole Miller for very useful discussions and comments and Collin Capano for a careful reading of the manuscript and comments. 
AB and PG acknowledge partial support from NASA Grant NNX12AN10G. PG was also supported during this work by an appointment to the NASA Postdoctoral Program at the Goddard Space Flight Center, administered by Oak Ridge Associated Universities through a contract with NASA. BSS acknowledges the support of the LIGO Visitor Program through the National Science Foundation award PHY-0757058, Max-Planck Institute of Gravitational Physics, Potsdam, Germany, and STFC grant ST/J000345/1. Results presented here were produced using the NEMO computing cluster at the Center for Gravitation and Cosmology at UWM under NSF Grants PHY-0923409 and PHY-0600953.

\appendix

\section{Summary of measurements}\label{sec:measurementSummary}

In this section we present tables of the $95\%$ credible intervals for five of the measured parameters: $\Mo$, $\eta$, $\Mc_{\rm obs}$, $d_L$, and $t_c$. Results are presented for two mass ratios ($q=1.25$ and $q=4$) and for two inclinations ($\theta_{\rm JN}=\pi/3$ and $\theta_{\rm JN}=0$). For all parameters except $t_c$, these values are scaled by their true values and converted into percentages. Uncertainty in $t_c$ is presented in ms. Table~\ref{tab:measurement_uncertainty_q125} presents values for $q=1.25$ and Table~\ref{tab:measurement_uncertainty_q4} presents results for $q=4$. For the two inclinations, results are side-by-side, with $\theta_{\rm JN}=0$ in parentheses.

\begin{center}
\begin{table*}
\begin{tabular*}{0.95\textwidth}{@{\extracolsep{\fill}}lccccccc}
\hline \hline
 SNR & $\Mo=50\Msun$ & $100\Msun$ & $150\Msun$ & $200\Msun$ & $300\Msun$ & $400\Msun$ & $500\Msun$\\
\hline
\multicolumn{8}{c}{$\Delta \Mo / \Mo$} \\
\hline
6 & 16.810 (16.658) & 6.895 (6.655) & 2.607 (2.815) & 0.612 (0.610) & 0.479 (0.515) & 0.472 (0.497) & 0.400 (0.410) \\
8 & 16.756 (16.464) & 0.159 (0.158) & 0.280 (0.285) & 0.288 (0.287) & 0.262 (0.251) & 0.239 (0.255) & 0.213 (0.214) \\
10 & 0.052 (0.049) & 0.086 (0.090) & 0.180 (0.178) & 0.202 (0.201) & 0.201 (0.200) & 0.180 (0.181) & 0.155 (0.159) \\
12 & 0.041 (0.040) & 0.065 (0.064) & 0.130 (0.130) & 0.166 (0.161) & 0.164 (0.169) & 0.149 (0.150) & 0.132 (0.125) \\
14 & 0.031 (0.033) & 0.050 (0.050) & 0.100 (0.102) & 0.137 (0.135) & 0.143 (0.139) & 0.128 (0.124) & 0.113 (0.105) \\
16 & 0.027 (0.028) & 0.042 (0.045) & 0.080 (0.082) & 0.116 (0.118) & 0.132 (0.129) & 0.109 (0.105) & 0.090 (0.088) \\
18 & 0.025 (0.025) & 0.038 (0.038) & 0.070 (0.074) & 0.100 (0.104) & 0.113 (0.107) & 0.097 (0.087) & 0.083 (0.075) \\
\hline
\multicolumn{8}{c}{$\Delta \eta / \eta$} \\
\hline
6 & 0.874 (0.872) & 0.866 (0.863) & 0.709 (0.725) & 0.488 (0.427) & 0.475 (0.484) & 0.512 (0.510) & 0.520 (0.524) \\
8 & 0.870 (0.871) & 0.201 (0.191) & 0.177 (0.181) & 0.182 (0.198) & 0.218 (0.227) & 0.221 (0.218) & 0.203 (0.186) \\
10 & 0.109 (0.104) & 0.135 (0.144) & 0.126 (0.130) & 0.132 (0.131) & 0.158 (0.146) & 0.136 (0.153) & 0.131 (0.097) \\
12 & 0.085 (0.082) & 0.114 (0.112) & 0.106 (0.099) & 0.113 (0.107) & 0.131 (0.129) & 0.120 (0.111) & 0.094 (0.075) \\
14 & 0.063 (0.067) & 0.086 (0.093) & 0.090 (0.086) & 0.089 (0.092) & 0.108 (0.100) & 0.100 (0.082) & 0.065 (0.053) \\
16 & 0.055 (0.057) & 0.077 (0.086) & 0.077 (0.073) & 0.080 (0.078) & 0.107 (0.096) & 0.082 (0.061) & 0.054 (0.038) \\
18 & 0.049 (0.050) & 0.071 (0.072) & 0.070 (0.069) & 0.072 (0.073) & 0.085 (0.074) & 0.069 (0.049) & 0.045 (0.034) \\
\hline
\multicolumn{8}{c}{$\Delta \Mc_{\rm obs} / \Mc_{\rm obs}$} \\
\hline
6 & 16.913 (16.613) & 6.594 (6.315) & 0.995 (1.293) & 0.605 (0.620) & 0.603 (0.602) & 0.626 (0.634) & 0.627 (0.614) \\
8 & 16.662 (16.350) & 0.169 (0.182) & 0.312 (0.324) & 0.333 (0.338) & 0.344 (0.344) & 0.347 (0.357) & 0.304 (0.292) \\
10 & 0.029 (0.027) & 0.091 (0.099) & 0.199 (0.200) & 0.237 (0.239) & 0.269 (0.263) & 0.240 (0.250) & 0.213 (0.197) \\
12 & 0.021 (0.021) & 0.065 (0.068) & 0.149 (0.146) & 0.203 (0.195) & 0.220 (0.223) & 0.203 (0.199) & 0.170 (0.150) \\
14 & 0.015 (0.016) & 0.049 (0.052) & 0.116 (0.118) & 0.163 (0.166) & 0.192 (0.190) & 0.178 (0.157) & 0.138 (0.123) \\
16 & 0.013 (0.013) & 0.043 (0.043) & 0.091 (0.095) & 0.140 (0.142) & 0.180 (0.173) & 0.146 (0.128) & 0.110 (0.098) \\
18 & 0.011 (0.012) & 0.037 (0.036) & 0.081 (0.084) & 0.125 (0.130) & 0.149 (0.137) & 0.127 (0.105) & 0.096 (0.086) \\
\hline
\multicolumn{8}{c}{$\Delta d_L / d_L$} \\
\hline
6 & 1.819 (2.438) & 1.252 (1.649) & 0.843 (1.078) & 0.629 (0.835) & 0.469 (0.655) & 0.449 (0.615) & 0.510 (0.658) \\
8 & 2.423 (3.066) & 1.470 (1.982) & 1.229 (1.569) & 0.873 (1.252) & 0.673 (0.932) & 0.687 (0.911) & 0.803 (1.090) \\
10 & 1.215 (1.616) & 1.211 (1.559) & 1.213 (1.628) & 1.179 (1.578) & 0.973 (1.319) & 0.973 (1.375) & 1.183 (1.560) \\
12 & 1.140 (1.513) & 1.084 (1.463) & 1.104 (1.526) & 1.133 (1.559) & 1.164 (1.576) & 1.270 (1.561) & 1.390 (1.461) \\
14 & 1.045 (1.333) & 0.980 (1.399) & 1.002 (1.417) & 1.030 (1.467) & 1.106 (1.404) & 1.234 (1.245) & 1.235 (1.009) \\
16 & 0.970 (1.418) & 0.896 (1.365) & 0.959 (1.430) & 1.004 (1.423) & 1.036 (1.307) & 1.035 (1.010) & 0.979 (0.821) \\
18 & 0.978 (1.386) & 0.895 (1.373) & 0.982 (1.378) & 0.991 (1.337) & 0.976 (1.118) & 0.923 (0.846) & 0.808 (0.688) \\
\hline
\multicolumn{8}{c}{$\Delta t_c$ (ms)} \\
\hline
6 & 187.8 (188.9) & 174.7 (173.3) & 48.3 (56.5) & 28.0 (26.2) & 34.8 (34.4) & 44.1 (44.4) & 46.1 (48.8) \\
8 & 191.3 (190.7) & 6.9 (6.8) & 11.3 (11.6) & 14.4 (14.8) & 20.9 (19.9) & 26.0 (26.2) & 30.6 (29.5) \\
10 & 3.5 (3.2) & 4.8 (4.8) & 8.1 (7.6) & 10.9 (10.5) & 16.3 (15.8) & 21.5 (20.2) & 24.0 (22.8) \\
12 & 3.0 (2.7) & 4.0 (3.9) & 6.3 (6.2) & 9.1 (8.9) & 13.8 (13.1) & 17.7 (16.3) & 19.7 (18.1) \\
14 & 2.5 (2.2) & 3.4 (3.1) & 5.1 (5.0) & 7.5 (7.3) & 11.5 (11.3) & 15.2 (13.6) & 16.4 (14.3) \\
16 & 2.2 (1.9) & 2.9 (2.6) & 4.3 (4.3) & 6.6 (6.6) & 10.8 (9.8) & 12.6 (11.1) & 14.1 (12.3) \\
18 & 2.0 (1.7) & 2.6 (2.5) & 3.8 (3.8) & 5.6 (5.8) & 9.1 (8.3) & 11.3 (9.6) & 12.2 (10.5) \\
\hline \hline
\end{tabular*}
\caption{Measurement uncertainties for \HM{} signals with \HM{} templates. The values are the (relative) widths of the $95\%$ credible intervals from the one-dimensional marginalized posterior distributions, scaled by the true value when indicated. For all runs, $q=1.25$ ($\eta=0.247$). $\theta_{\rm JN}=\pi/3$ and $\theta_{\rm JN}=0$ results are shown side-by-side, with the latter in parentheses.}
\label{tab:measurement_uncertainty_q125}
\end{table*}

\begin{table*}
\begin{tabular*}{0.95\textwidth}{@{\extracolsep{\fill}}lccccccc}
\hline \hline
 SNR & $\Mo=50\Msun$ & $100\Msun$ & $150\Msun$ & $200\Msun$ & $300\Msun$ & $400\Msun$ & $500\Msun$\\
\hline
\multicolumn{8}{c}{$\Delta \Mo / \Mo$} \\
\hline
6 & 16.206 (17.067) & 7.659 (7.722) & 3.378 (3.896) & 1.821 (2.195) & 0.966 (1.091) & 0.822 (0.930) & 0.586 (0.690) \\
8 & 16.749 (16.519) & 0.336 (0.452) & 0.523 (0.626) & 0.417 (0.419) & 0.362 (0.354) & 0.334 (0.320) & 0.324 (0.314) \\
10 & 15.992 (0.047) & 0.138 (0.136) & 0.298 (0.229) & 0.310 (0.260) & 0.292 (0.258) & 0.269 (0.262) & 0.263 (0.244) \\
12 & 0.034 (0.033) & 0.111 (0.102) & 0.130 (0.110) & 0.233 (0.175) & 0.253 (0.212) & 0.242 (0.214) & 0.226 (0.203) \\
14 & 0.028 (0.025) & 0.091 (0.082) & 0.098 (0.084) & 0.176 (0.115) & 0.220 (0.170) & 0.212 (0.174) & 0.182 (0.167) \\
16 & 0.021 (0.021) & 0.078 (0.068) & 0.084 (0.068) & 0.111 (0.081) & 0.196 (0.147) & 0.185 (0.153) & 0.172 (0.144) \\
18 & 0.218 (0.044) & 0.072 (0.061) & 0.075 (0.058) & 0.085 (0.065) & 0.173 (0.124) & 0.163 (0.128) & 0.142 (0.122) \\
\hline
\multicolumn{8}{c}{$\Delta \eta / \eta$} \\
\hline
6 & 1.347 (1.348) & 1.343 (1.342) & 1.290 (1.314) & 1.140 (1.233) & 1.035 (1.066) & 1.063 (1.043) & 1.021 (1.021) \\
8 & 1.345 (1.349) & 0.817 (0.824) & 0.785 (0.826) & 0.753 (0.771) & 0.717 (0.776) & 0.693 (0.744) & 0.720 (0.775) \\
10 & 1.346 (0.126) & 0.478 (0.443) & 0.714 (0.623) & 0.705 (0.650) & 0.648 (0.627) & 0.666 (0.645) & 0.641 (0.645) \\
12 & 0.082 (0.079) & 0.344 (0.308) & 0.455 (0.389) & 0.595 (0.490) & 0.597 (0.539) & 0.594 (0.551) & 0.590 (0.553) \\
14 & 0.062 (0.057) & 0.272 (0.245) & 0.348 (0.295) & 0.492 (0.354) & 0.547 (0.455) & 0.541 (0.459) & 0.497 (0.473) \\
16 & 0.047 (0.047) & 0.227 (0.196) & 0.288 (0.237) & 0.369 (0.279) & 0.503 (0.384) & 0.502 (0.412) & 0.468 (0.408) \\
18 & 0.223 (0.123) & 0.203 (0.170) & 0.255 (0.202) & 0.281 (0.226) & 0.450 (0.340) & 0.445 (0.354) & 0.415 (0.358) \\
\hline
\multicolumn{8}{c}{$\Delta \Mc_{\rm obs} / \Mc_{\rm obs}$} \\
\hline
6 & 20.744 (22.182) & 9.828 (10.135) & 3.308 (4.837) & 1.480 (2.081) & 1.268 (1.391) & 1.276 (1.368) & 1.131 (1.119) \\
8 & 21.574 (21.289) & 0.613 (0.780) & 0.917 (0.934) & 0.822 (0.793) & 0.780 (0.761) & 0.763 (0.745) & 0.748 (0.762) \\
10 & 20.421 (0.027) & 0.164 (0.158) & 0.640 (0.521) & 0.691 (0.595) & 0.682 (0.628) & 0.666 (0.639) & 0.644 (0.623) \\
12 & 0.016 (0.016) & 0.109 (0.098) & 0.263 (0.208) & 0.551 (0.420) & 0.609 (0.517) & 0.597 (0.534) & 0.579 (0.528) \\
14 & 0.011 (0.011) & 0.080 (0.072) & 0.150 (0.130) & 0.428 (0.274) & 0.547 (0.430) & 0.545 (0.443) & 0.476 (0.446) \\
16 & 0.008 (0.009) & 0.064 (0.057) & 0.114 (0.101) & 0.267 (0.194) & 0.485 (0.366) & 0.479 (0.392) & 0.450 (0.378) \\
18 & 0.208 (0.033) & 0.056 (0.049) & 0.098 (0.084) & 0.179 (0.151) & 0.432 (0.316) & 0.424 (0.330) & 0.379 (0.330) \\
\hline
\multicolumn{8}{c}{$\Delta d_L / d_L$} \\
\hline
6 & 2.340 (3.107) & 1.792 (2.252) & 1.503 (2.007) & 1.259 (1.658) & 1.061 (1.467) & 1.119 (1.433) & 1.252 (1.537) \\
8 & 3.197 (4.272) & 2.847 (3.761) & 2.465 (3.254) & 2.017 (2.683) & 1.631 (2.152) & 1.693 (2.014) & 1.909 (2.189) \\
10 & 3.869 (1.842) & 1.226 (1.453) & 2.090 (2.034) & 1.989 (2.128) & 1.717 (1.780) & 1.691 (1.534) & 1.740 (1.495) \\
12 & 1.144 (1.388) & 0.926 (1.014) & 1.019 (1.048) & 1.330 (1.271) & 1.338 (1.256) & 1.272 (1.127) & 1.263 (1.039) \\
14 & 0.988 (1.147) & 0.785 (0.875) & 0.750 (0.821) & 0.995 (0.941) & 1.083 (0.940) & 1.006 (0.889) & 0.926 (0.839) \\
16 & 0.878 (1.036) & 0.688 (0.765) & 0.656 (0.693) & 0.747 (0.719) & 0.924 (0.806) & 0.852 (0.743) & 0.803 (0.676) \\
18 & 2.011 (1.183) & 0.609 (0.670) & 0.609 (0.615) & 0.619 (0.634) & 0.795 (0.668) & 0.750 (0.644) & 0.657 (0.589) \\
\hline
\multicolumn{8}{c}{$\Delta t_c$ (ms)} \\
\hline
6 & 189.9 (190.2) & 185.6 (185.2) & 142.4 (166.4) & 58.8 (100.6) & 54.6 (57.4) & 65.2 (65.3) & 71.4 (66.7) \\
8 & 191.5 (188.9) & 11.0 (11.9) & 17.3 (19.6) & 19.2 (19.3) & 26.8 (25.8) & 36.7 (34.2) & 44.6 (44.7) \\
10 & 188.8 (4.4) & 7.0 (6.7) & 11.1 (10.0) & 14.9 (13.1) & 21.3 (19.0) & 26.1 (24.2) & 32.3 (28.1) \\
12 & 3.6 (3.4) & 6.0 (5.3) & 8.2 (7.0) & 11.8 (9.7) & 18.0 (15.5) & 22.3 (19.3) & 24.7 (21.2) \\
14 & 3.1 (2.9) & 5.1 (4.5) & 6.7 (5.7) & 9.4 (7.8) & 15.0 (12.0) & 19.1 (15.6) & 19.7 (16.7) \\
16 & 2.9 (2.6) & 4.5 (3.9) & 5.8 (4.8) & 7.6 (6.1) & 13.1 (10.6) & 15.8 (13.1) & 17.6 (13.6) \\
18 & 9.0 (5.1) & 4.0 (3.4) & 5.1 (4.3) & 6.5 (5.4) & 11.9 (8.7) & 13.6 (10.6) & 14.4 (11.6) \\
\hline \hline
\end{tabular*}
\caption{Measurement uncertainties for \HM{} signals with \HM{} templates. The values are the (relative) widths of the $95\%$ credible intervals from the one-dimensional marginalized posterior distributions, scaled by the true value when indicated. For all runs, $q=4$ ($\eta=0.16$). $\theta_{\rm JN}=\pi/3$ and $\theta_{\rm JN}=0$ results are shown side-by-side, with the latter in parentheses.}
\label{tab:measurement_uncertainty_q4}
\end{table*}
\end{center}

\bibliography{refs}

\end{document}